\documentclass{article}
\usepackage{tcolorbox}
\usepackage{subcaption}
\usepackage{authblk}
\usepackage[english]{babel}

\usepackage[letterpaper,top=2cm,bottom=2cm,left=2cm,right=2cm,marginparwidth=1cm]{geometry}

\usepackage{algorithm}
\usepackage{algpseudocode}
\usepackage{amsmath}

\usepackage[utf8]{inputenc}
\usepackage{graphicx}%
\usepackage{multirow}%
\usepackage{amsmath,amssymb,amsfonts}%
\usepackage{amsthm}%
\usepackage{mathrsfs}%
\usepackage[title]{appendix}%
\usepackage{xcolor}%
\usepackage{textcomp}%
\usepackage{manyfoot}%
\usepackage{booktabs}%
\usepackage{siunitx}
\usepackage{pifont}
\newcommand{\xmark}{\ding{55}}

\usepackage{booktabs}

\usepackage{enumitem}
\usepackage{algorithm}%
\usepackage{algorithmicx}%
\usepackage{algpseudocode}%
\usepackage{listings}%
\usepackage{graphicx}
\usepackage{longtable}
\usepackage{array}
\usepackage{booktabs}
\usepackage{multirow}
\usepackage{adjustbox}
\usepackage{xcolor}
\usepackage{url}
\usepackage{hyperref}

\definecolor{codegreen}{rgb}{0,0.6,0}
\definecolor{codegray}{rgb}{0.5,0.5,0.5}
\definecolor{codepurple}{rgb}{0.58,0,0.82}
\definecolor{backcolour}{rgb}{0.95,0.95,0.92}

\lstdefinestyle{mystyle}{
    backgroundcolor=\color{backcolour},   
    commentstyle=\color{codegreen},
    keywordstyle=\color{magenta},
    numberstyle=\tiny\color{codegray},
    stringstyle=\color{codepurple},
    basicstyle=\ttfamily\footnotesize,
    breakatwhitespace=false,         
    breaklines=true,                 
    captionpos=b,                    
    keepspaces=true,                 
    numbers=left,                    
    numbersep=5pt,                  
    showspaces=false,                
    showstringspaces=false,
    showtabs=false,                  
    tabsize=2
}

\usepackage{tcolorbox}       
\tcbuselibrary{skins}
\tcbuselibrary{breakable}    

\newtcolorbox{emserqbox}[2][]{%
  enhanced,
  breakable,
  colback=gray!8,
  colframe=gray!50,
  boxrule=0.4pt,
  arc=2pt,
  left=7pt,right=7pt,
  top=6pt,bottom=6pt,
  fonttitle=\bfseries,
  coltitle=black,
  title={Answer to RQ#2},
  #1
}

\newenvironment{rqanswer}[1]{%
  \begin{emserqbox}{#1}%
}{%
  \end{emserqbox}%
}

\usepackage{fancyhdr}
\title{Characterizing Bugs and Quality Attributes in Quantum Software}

\title{Characterizing Bugs and Quality Attributes in Quantum Software: A Large-Scale Empirical Study}

\date{\vspace{-1em}\small\scshape A Preprint}

\author[1]{Mir Mohammad Yousuf\thanks{Corresponding author. Email: yousuf\_2022phaite006@nitsri.ac.in}}
\author[2]{Shabir Ahmad Sofi}

\affil[1,2]{Department of Information Technology\\
NIT Srinagar, J\&K, India\\
\href{mailto:yousuf_2022phaite006@nitsri.ac.in} {\texttt{yousuf\_2022phaite006@nitsri.ac.in}} and 
\href{mailto:shabir@nitsri.ac.in}{\texttt{shabir@nitsri.ac.in}}}

\begin{document}
\pagestyle{plain}
\maketitle
\abstract{

Quantum Software Engineering (QSE) is essential for ensuring the reliability and maintainability of hybrid quantum-classical systems, yet empirical evidence on how bugs emerge and affect quality in real-world quantum projects remains limited. This study presents the first ecosystem-scale longitudinal analysis of software bugs across 123 open source quantum repositories from 2012 to 2024, spanning eight functional categories, including full-stack libraries, simulators, annealing, algorithms, compilers, assembly, cryptography, and experimental computing. Using a mixed method approach combining repository mining, static code analysis, issue metadata extraction, and a validated rule-based classification framework, we analyze 32,296 verified bug reports. Results show that full-stack libraries and compilers are the most bug-prone categories due to circuit, gate, and transpilation-related issues, while simulators are mainly affected by measurement and noise modeling errors. Classical bugs primarily impact usability and interoperability, whereas quantum-specific bugs disproportionately degrade performance, maintainability, and reliability. Longitudinal analysis indicates ecosystem maturation, with bug densities peaking between 2017 and 2021 and declining thereafter. High-severity bugs cluster in cryptography, experimental computing, and compiler toolchains. Repositories employing automated testing detect more bugs and resolve issues faster. A negative binomial regression further shows that automated testing is associated with an approximate 60 percent reduction in expected bug incidence. Overall, this work provides the first large-scale data-driven characterization of quantum software bugs and offers empirical guidance for improving testing, documentation, and maintainability practices in QSE.
}

\section{Introduction}
\label{intro}

Quantum computing has emerged as a transformative computational paradigm, promising exponential speedups in domains such as optimization, simulation, and cryptography~\cite{feynman2018simulating,preskill2018quantum}.  
As foundational studies and manifestos argue, the growing complexity of quantum platforms demands the establishment of Quantum Software Engineering (QSE) as a systematic discipline grounded in modernization, quality management, and structured engineering processes~\cite{piattini2021toward,piattini2020talavera,perez2021software}.
However, developing reliable quantum software remains a major engineering challenge due to its hybrid architecture, which integrates classical control logic with quantum circuits.  
This hybrid design introduces complex defect behaviors, ranging from quantum-specific issues (e.g., decoherence, gate fidelity errors, and transpilation failures) to classical ones (e.g., syntax, integration, and usability bugs), complicating testing and debugging~\cite{mandal2025quantum,montangero2023programming,ramalho2024testing}.

Quantum programs differ from classical systems not only in execution semantics but also in observability. Their probabilistic outputs and limited internal visibility restrict conventional debugging and verification approaches.  
Compounding this challenge, the current ecosystem lacks mature abstractions, reusable components, and standardized quality assessment practices, as highlighted by recent systematic evaluations of quantum software platforms~\cite{serrano2022quantum}.  
The fragmentation of toolchains, languages, and backend interfaces further limits the adoption of automated testing, static analysis, and continuous integration.  
As a result, understanding how defects manifest, evolve, and affect software quality is central to advancing QSE and establishing engineering foundations for trustworthy quantum applications.

\medskip
\noindent
The quantum software landscape has expanded rapidly with the rise of platforms such as \textit{Qiskit}, \textit{Cirq}, \textit{PennyLane}, and \textit{QuTiP}.  
Existing research has provided important but fragmented insights into correctness~\cite{adedoyin2018quantum}, design and architecture~\cite{barbosa2020software}, and early debugging practices~\cite{paltenghi2022bugs}.  
Benchmark datasets such as \textit{Bugs4Q}~\cite{bugs4q} and \textit{QBugs}~\cite{qbugs} have enabled controlled studies, while recent surveys~\cite{quetschlich2025experience,zappin2025challenges,paltenghi2024survey} have cataloged open challenges in testing and tooling.  
However, these efforts are limited in scope, typically focusing on individual frameworks, narrow time windows, or qualitative analyses.  
They do not capture the \textit{longitudinal evolution of defects, severity distributions, and their implications for software quality} across the broader ecosystem.  
Longitudinal evidence is essential to reveal trends in testing maturity, maintainability, and defect resolution practices, thereby providing a data-driven perspective on the evolution and stability of the quantum software ecosystem.

\medskip
\noindent
To fill this gap, we present a large-scale, empirical study of \textbf{123 open-source quantum software repositories }comprising 32,296 bug reports.  
Our hybrid methodology integrates static code analysis, repository mining, issue metadata extraction, and rule-based defect classification.  
Each defect is characterized by its \textit{type} (classical or quantum-specific), \textit{category}, \textit{severity}, \textit{affected quality attributes}, and, when applicable, \textit{quantum-specific subtype}.  
We further examine how repository-level characteristics, programming language, documentation quality, testing framework adoption, and code complexity, influence defect detection, reporting, and resolution.

\subsubsection*{Empirical Focus:}
This study adopts a multi-dimensional perspective on defect behavior, addressing six interconnected facets of software quality:
\begin{itemize}
    \item \textbf{Temporal evolution:} how bug counts and categories have evolved over time, reflecting ecosystem maturity.
    \item \textbf{Severity dynamics:} how bug severity varies by repository type and its implications for debugging and reliability.
    \item \textbf{Quality attributes:} how bugs affect maintainability, reliability, performance, and usability.
    \item \textbf{Development practices:} how programming languages and documentation quality shape usability and adoption;
    \item \textbf{Testing and complexity:} how testing frameworks, coverage tools, and code complexity influence bug discovery and resolution.
    \item \textbf{Quantum-specific defects:}  how quantum circuits, gates, and transpilation errors are distributed across repository categories.
\end{itemize}

\subsubsection*{Research Questions:}

To guide this empirical investigation, we formulate twelve research questions (RQs) grouped into six thematic clusters, reflecting the multifaceted nature of quality, reliability, and defect behavior in quantum software systems.

\begin{enumerate}
    \item \textbf{Bug Evolution and Distribution}
    \begin{itemize}
        \item \textbf{RQ1:} How have bug counts and categories in open-source quantum software evolved over time, and how do these trends differ across repository types?
    \end{itemize}

    \item \textbf{Bug Severity and Debugging}
    \begin{itemize}
        \item \textbf{RQ2:} How does bug severity vary across different categories of quantum software repositories, and what implications does this have for debugging and reliability in quantum computing?
    \end{itemize}

    \item \textbf{Quality Attributes and Reliability}
    \begin{itemize}
        \item \textbf{RQ3:} How do different types of quantum software repositories vary in terms of quality attributes affected by bugs, and what insights can be derived to improve software reliability, maintainability, and usability?
        \item \textbf{RQ4:} How do classical and quantum-specific bugs differ in their impact on software quality attributes?
        \item \textbf{RQ5:} How do different bug categories (e.g., compatibility, functional, quantum-specific) impact key quality attributes, and what unique challenges do quantum-specific issues present?
    \end{itemize}

    \item \textbf{Programming Languages and Documentation}
    \begin{itemize}
        \item \textbf{RQ6:} How does the distribution of programming languages influence development, performance, and usability of quantum software?
        \item \textbf{RQ7:} How does documentation quality (README, source docs, tutorials) impact usability and adoption?
    \end{itemize}

    \item \textbf{Testing and Complexity}
    \begin{itemize}
        \item \textbf{RQ8:} How do testing frameworks and coverage tools affect bug detection and reporting?
        \item \textbf{RQ9:} How do cyclomatic complexity, testing practices, and coverage adoption vary across languages, and what are the implications for reliability?
        \item \textbf{RQ10:} What is the impact of automated testing on code quality and bug detection in quantum software repositories?
    \end{itemize}

    \item \textbf{Quantum-Specific Bugs}
    \begin{itemize}
        \item \textbf{RQ11:} Which types of quantum-specific bugs are most prevalent, and what insights can be derived from their distribution?
        \item \textbf{RQ12:} How do quantum-specific bug distributions vary across repository categories, and what do these variations imply for QSE practice?
    \end{itemize}
\end{enumerate}

\noindent
Table~\ref{tab:rq_overview} summarizes the twelve research questions investigated in this study,
along with their analytical focus, key variables, and empirical basis.
The table serves as a bridge between the conceptual framing of the RQs
and the methodological procedures detailed in Section~\ref{sec:methodology}.
Each question targets a distinct but complementary aspect of software quality and reliability,
ranging from temporal defect evolution and severity patterns to testing practices and quantum-specific issues.

\begin{table*}[ht]
\centering
\small
\caption{Overview of Research Questions, Analytical Focus, and Empirical Basis}
\label{tab:rq_overview}
\begin{tabular}{p{0.8cm}p{4cm}p{5cm}p{5cm}}
\hline
\textbf{RQ} & \textbf{Analytical Focus} & \textbf{Key Variables / Dimensions} & \textbf{Empirical Basis (Data \& Methods)} \\
\hline
\textbf{RQ1} & Temporal evolution of defects across repositories & Bug counts, categories, repository type, time (2012–2024) & Longitudinal mining of 123 repositories; issue metadata extraction; temporal trend analysis \\
\textbf{RQ2} & Variation of bug severity across repository types & Severity levels (low–critical); repository category & Descriptive severity profiling; cross-category comparison; visualization analysis \\
\textbf{RQ3} & Quality attributes affected by bugs across repository types & Attributes: usability, maintainability, performance, interoperability, reliability & Bug-to-quality mapping; attribute frequency analysis \\
\textbf{RQ4} & Differential impact of classical vs.\ quantum-specific defects & Defect type (classical vs.\ quantum-specific); quality impact & Comparative attribute-impact analysis; defect-type classification \\
\textbf{RQ5} & Effects of bug categories on quality attributes & Bug category (compatibility, functional, quantum-specific, etc.) & Category-wise mapping; statistical aggregation \\
\textbf{RQ6} & Influence of programming language on development, performance, and usability & Primary programming language; repository characteristics & Language identification; correlation analysis with bug and complexity metrics \\
\textbf{RQ7} & Effect of documentation quality on usability and adoption & README quality; inline docs; tutorials; user engagement & Documentation scoring; contributor-activity correlation \\
\textbf{RQ8} & Impact of testing frameworks and coverage tools on defect visibility & Test presence; coverage-tool adoption; bug reporting rate & Testing metadata extraction; bug-count comparison; descriptive and regression analysis \\
\textbf{RQ9} & Relationship between complexity, testing, and reliability & Cyclomatic complexity; testing adoption; coverage ratio & Static code analysis; coverage metrics; multivariate regression \\
\textbf{RQ10} & Impact of automated testing on code quality and defect incidence & Automated testing presence; bug incidence; static code quality & Regression modeling (IRR estimation); controls for size and complexity \\
\textbf{RQ11} & Prevalence of quantum-specific bug types & Quantum defect subtypes (circuit, gate, transpilation, noise, etc.) & Quantum-specific classification; frequency analysis; visualization \\
\textbf{RQ12} & Distribution of quantum-specific bugs across repository categories & Defect subtype × repository category & Cross-tabulation; domain-level interpretation for QSE practice \\
\hline
\end{tabular}
\end{table*}

\subsubsection*{Contributions:}
This work makes the following key contributions:

\begin{enumerate}
    \item \textbf{A longitudinal, ecosystem-scale analysis of quantum software defects:}  
    We conduct the largest empirical study to date of quantum software repositories, analyzing 123 open-source projects over twelve years (2012--2024) to characterize ecosystem evolution, defect trends, and maturity across frameworks, compilers, simulators, QML, annealing tools, and domain libraries.

    \item \textbf{A comprehensive classification of bug types, severities, and quality-attribute impacts:}  
    We provide fine-grained categorizations of classical vs.\ quantum-specific defects, analyze severity distributions, and map defects to affected quality attributes such as usability, maintainability, performance, interoperability, and correctness.

    \item \textbf{A detailed examination of testing, complexity, and automated quality practices in QSE:}  
    We evaluate testing adoption, coverage integration, cyclomatic complexity, static analysis, and CI usage across languages (Python, C/C++, Julia, OpenQASM) and use NB-GLM regression to quantify how these factors influence defect visibility and maintainability.

    \item \textbf{The first cross-cutting analysis of quantum-specific defects across circuit, gate, transpilation, and hybrid-interface layers:}  
    We identify defect hotspots, variation across repository categories, and impacts on correctness and performance, extending prior taxonomies with evidence-backed insights from compilers, simulators, and orchestration layers.

    \item \textbf{Empirically grounded implications and practitioner-oriented recommendations for QSE:}  
    We derive actionable guidance for debugging, documentation, testing, tooling, classical--quantum integration, and quantum-specific verification, offering a practical roadmap for improving reliability and maintainability in quantum software.
\end{enumerate}

\medskip
\noindent
The remainder of this paper is organized as follows:  
Section~\ref{related_work} reviews prior research in software bug classification and quantum software engineering.  
Section~\ref{sec:methodology} details our methodology for data collection, static analysis, and rule-based defect classification.  
Section~\ref{sec:resultsandobservations} presents empirical findings across all twelve research questions.  
Section~\ref{sec:discussion} synthesizes the results and provides a discussion..  
Sections~\ref{sec:conclusion} and~\ref{sec:future_directions} summarize key insights and outline future directions for advancing the reliability and maintainability of quantum software systems.Finally Section~\ref{sec:threats_validity} discusses limitations and validity threats.

\section{Related Work}
\label{related_work}

QSE has emerged as a discipline aimed at establishing systematic processes, methods, and quality principles for quantum software development.  
Conceptual foundations, including the Talavera Manifesto and studies ~\cite{piattini2021toward,piattini2020talavera,perez2021software}, emphasize the need for modernization, structured development lifecycles, and engineering best practices to support the rapid evolution of quantum technologies.  
Despite these calls, the ecosystem remains immature in terms of abstractions, platform heterogeneity, and component modularity, as documented in recent assessments of quantum software platforms~\cite{serrano2022quantum}.  
Existing empirical efforts, though valuable, have primarily focused on narrow domains, small datasets, or controlled benchmarks.  
Our study contributes the first ecosystem-scale, longitudinal analysis of 32,296 issues across 123 repositories spanning 2012–2024.

\subsection*{Benchmarking and Controlled Testing Studies}
Benchmark efforts such as QBugs~\cite{qbugs} and Bugs4Q~\cite{bugs4q} provide reproducible datasets for tool validation but remain limited in temporal and ecosystem scope.  
The broader QSE literature emphasizes that controlled environments alone cannot capture the architectural, language-level, and process-related quality challenges seen in real-world codebases~\cite{piattini2021toward,piattini2020talavera}.  
These insights underscore the need for large-scale repository-based studies that reflect natural defect evolution across frameworks and application domains.

\subsection*{Empirical Analyses of Quantum Bugs}
Prior empirical studies have examined quantum-specific defects at smaller scales~\cite{paltenghi2022bugs,metwalli2023categorization,zhao2023empirical,luo2022comprehensive}.  
While these provide valuable taxonomies, they are limited by dataset size, scope, and lack of temporal breadth.  
Recent platform assessments~\cite{serrano2022quantum} further highlight structural issues such as platform fragmentation, lack of reusable components, and immature toolchains, gaps that our large-scale longitudinal analysis helps contextualize.

\subsection*{Experience-Based and Practitioner Studies}
Practitioner-focused studies reveal challenges in debugging, tool immaturity, and ecosystem fragmentation~\cite{quetschlich2025experience,zappin2025challenges}.  
These perspectives align with platform-level quality assessments~\cite{serrano2022quantum} and conceptual QSE frameworks~\cite{piattini2021toward}.  
Our work bridges these qualitative insights with ecosystem-scale empirical evidence, enabling quantitative analysis of how testing, documentation, and architectural complexity shape reliability across QSE.

\begin{table}[h!]
\centering
\caption{Representative Studies on Quantum Software Bugs}
\renewcommand{\arraystretch}{1.25}
\small
\begin{tabular}{p{3.5cm}p{3cm}p{4cm}p{5cm}}
\hline
\textbf{Study} & \textbf{Scope} & \textbf{Key Findings} & \textbf{Limitations} \\ \hline

\textbf{QBugs}~(2021)~\cite{qbugs} & Benchmark of reproducible bugs & Provides curated dataset and infrastructure for debugging experiments. & Focused on algorithmic benchmarks; lacks ecosystem diversity and temporal context. \\ \hline

\textbf{Bugs4Q}~(2023)~\cite{bugs4q} & Benchmark of real-world bugs & Enables reproducible and comparative testing studies. & Limited to controlled benchmark setting; no longitudinal or ecosystem coverage. \\ \hline

\textbf{Paltenghi \& Pradel}~(2022)~\cite{paltenghi2022bugs} & 223 bugs from 18 platforms & 39.9\% quantum-specific; proposed ten bug patterns. & Limited platform scope; static dataset. \\ \hline

\textbf{Metwalli \& Van Meter}~(2023)~\cite{metwalli2023categorization} & Forum and repository data & Four major categories: job handling, post-processing, semantics, and circuits. & Qualitative data; limited generalizability. \\ \hline

\textbf{Zhao et al.}~(2023)~\cite{zhao2023empirical} & 391 bugs from 22 QML frameworks & 28\% quantum-specific; highlighted device inconsistency and qubit manipulation errors. & Narrow domain; dependency-related complexity. \\ \hline

\textbf{Luo et al.}~(2022)~\cite{luo2022comprehensive} & 96 bug fixes across 4 languages & 80\% quantum-specific; emphasized debugging tool need. & Small dataset; limited language diversity. \\ \hline

\textbf{Quetschlich \& Di Matteo}~(2025)~\cite{quetschlich2025experience} & Developer experience and code review & Introduced taxonomy of recurring quantum bug types. & Qualitative; not validated on large datasets. \\ \hline

\textbf{This Study} & 123 repositories (2012–2024); 32,296 issues & Ecosystem-scale, rule-based classification by type, severity, and quality impact; integrates testing, documentation, and complexity metrics over time. & Broader in scope; some context-dependent defects may remain unclassified. \\ \hline
\end{tabular}
\label{tab:related_work}
\end{table}

Unlike prior benchmark, qualitative, or domain-specific efforts, our work provides a unified, ecosystem-wide characterization of quantum software defects across a twelve-year period.  
By combining rule-based classification~\cite{yousuf2026bug} with static analysis and longitudinal repository mining, it bridges the gap between controlled experimentation and real-world development practice.  
This integration offers the first comprehensive empirical foundation for understanding how testing maturity, documentation, and architectural complexity co-evolve to shape reliability across the quantum software ecosystem.

\section{Methodology}
\label{sec:methodology}

\begin{figure}[h!]
\centering
\includegraphics[width=1.05\linewidth]{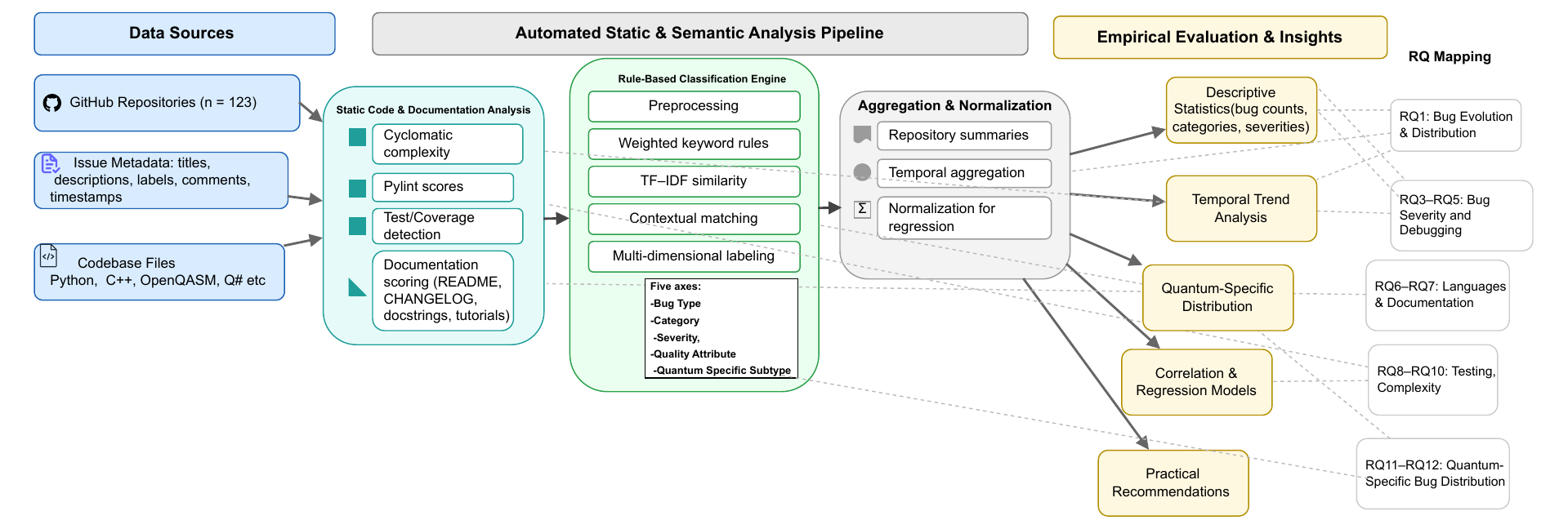}
\caption{Empirical framework illustrating repository selection, issue extraction, rule-based classification, static analysis, aggregation, and RQ mapping.}
\label{fig:empirical_framework}
\end{figure}

To investigate software defects in quantum computing systems, we conducted a large-scale empirical study of 123 open-source quantum software repositories spanning the period 2012–2024.  
Our methodology integrates repository mining, static code analysis, and a validated rule-based classification framework to examine bugs across multiple dimensions, including type, category, severity, quality impact, and temporal evolution.  
The complete workflow, including data sources, analytical stages, and research question mapping, is summarized in Figure~\ref{fig:empirical_framework}.  
This figure provides a unified view of the empirical pipeline, linking repository selection, issue extraction, rule-based classification, static and documentation analysis, and statistical modeling.

\begin{enumerate}
    \item \textbf{Repository Selection and Categorization:}  
    Repositories were selected from GitHub using the curated listings provided by the Quantum Open Source Foundation (QOSF)~\cite{qosf_project_list}.  
    Projects were classified into eight functional domains representing the spectrum of the quantum software ecosystem.

    \medskip
    \noindent\textbf{Repository Categories:}  
    To ensure clarity and consistency, repositories were grouped into eight categories based on their functional scope:
    \begin{itemize}
        \item \textit{Quantum Full Stack Libraries:} End-to-end SDKs integrating circuit design, simulation, and execution (e.g., Qiskit, Cirq, PennyLane).
        \item \textit{Quantum Compilers:} Tools that optimize and transpile circuits for hardware execution.
        \item \textit{Quantum Simulators:} Frameworks for classical simulation of quantum circuits and noise models;
        \item \textit{Quantum Algorithms:} Repositories implementing algorithms for optimization, chemistry, or QML.
        \item \textit{Quantum Assembly:} Low-level language toolchains or transpilers for hardware interfacing.
        \item \textit{Quantum Annealing :} Software focused on quantum annealing and Ising-model optimization.
        \item \textit{Quantum Experimental Computing :} Systems interfacing directly with experimental or hardware control environments.
        \item \textit{Quantum Cryptography :} Security, encryption, and communication-focused frameworks.  
    \end{itemize}

    This taxonomy ensures balanced coverage of both hardware-oriented and algorithmic repositories, consistent with prior QSE classifications~\cite{paltenghi2022bugs,luo2022comprehensive,quetschlich2025experience}.

    \item \textbf{Bug Metadata Extraction:}  
    Issue data were collected using the GitHub REST API, capturing issue identifiers, titles, descriptions, labels, timestamps, and associated comments. 

    \item \textbf{Rule-Based Bug Classification:}  
    We applied a validated rule-based classification framework~\cite{yousuf2026bug}, designed to systematically capture both classical and quantum-domain defects across multiple analytical dimensions.  
    Each issue was labeled using weighted keyword rules, TF--IDF similarity, and contextual pattern matching along five axes:

    \begin{enumerate}[label=(\alph*)]
        \item \textbf{Bug Type:} Classifies each defect as \textit{Quantum}, \textit{Classical}, or \textit{Uncategorized}, depending on whether it originates from quantum operations or classical control logic.  
        \item \textbf{Bug Category:} Denotes the underlying defect nature, \textit{Functional}, \textit{Logical}, \textit{Syntax}, \textit{Compatibility}, or related classes.  
        \item \textbf{Severity:} Reflects defect impact, categorized as \textit{Low}, \textit{Medium}, \textit{High}, or \textit{Critical}.  
        \item \textbf{Quality Attribute:} Maps each defect to the most affected ISO/IEC 25010 quality dimension~\cite{ISO25010}, such as \textit{Reliability}, \textit{Maintainability}, or \textit{Usability}.  
        \item \textbf{Quantum-Specific Subtype:} Refines quantum-domain bugs into \textit{Circuit}, \textit{Gate}, \textit{Measurement}, \textit{Noise}, \textit{Transpilation}, or \textit{Hardware Interaction} issues.  
    \end{enumerate}

    The framework was originally introduced and validated by Yousuf et al.~\cite{yousuf2026bug}, achieving robust accuracy (0.81–0.85) and macro–F1 scores (0.66–0.77) against manually annotated datasets.  
    Its multi-dimensional design enables consistent, scalable classification across heterogeneous quantum software ecosystems.  
    Further details on validation, reproducibility checks, and reliability verification are provided in Section~\ref{validation}.

    \item \textbf{Static Code and Repository Analysis:}  
    Following classification, we performed repository-level static analysis and documentation scoring to examine how code complexity, testing practices, and documentation quality relate to defect characteristics.  
    For each repository, we extracted static and structural metrics to assess code quality, maintainability, and documentation practices:
    \begin{enumerate}[label=(\alph*)]
        \item \textbf{Cyclomatic complexity}, computed using \texttt{radon}~\cite{radon2025}, quantified structural complexity and maintainability~\cite{kumar2023formalization,ebert2016cyclomatic}.  
        \item \textbf{Pylint scores} captured overall code quality and adherence to Pythonic conventions~\cite{pylint2025,yousuf2022analysis}.  
        \item \textbf{Testing analysis} detected the presence of unit tests (e.g., \texttt{test\_*.py}) and coverage tools (e.g., \texttt{coverage.py}).  
        \item \textbf{Documentation scoring} evaluated the completeness of essential artifacts (\texttt{README}, \texttt{CHANGELOG}, user guides, tutorials).  
    \end{enumerate}
    These metrics formed the quantitative foundation for subsequent correlation and regression analyses.

    \item \textbf{Data Aggregation and Statistical Analysis:}  
    Repository- and issue-level data were aggregated to support descriptive, comparative, and inferential analysis.  
    Aggregation included normalization by repository size and issue volume, followed by computation of summary statistics (means, medians, and distributions).  
    Inferential analyses employed correlation and regression modeling to assess relationships between testing practices, documentation quality, code complexity, and defect characteristics.  
    Longitudinal analyses captured ecosystem evolution and temporal maturity trends.
\end{enumerate}

This integrated workflow provides a reproducible and transparent foundation for analyzing defect patterns, software quality attributes, and testing practices across the quantum software ecosystem.  
All scripts, classification rules, and analysis configurations were implemented in Python and will be made available upon publication to ensure full reproducibility of results.

\subsection{Data Collection}
We curated a dataset of 123 open-source quantum computing repositories from GitHub, covering a diverse range of frameworks, programming languages, and application domains.
Table~\ref{tab:repo_summary} summarizes the number of repositories, cumulative bug counts, and mean repository sizes across categories.

The selected repositories span a broad spectrum of quantum software domains, including Full Stack Libraries, Compilers, Simulators, Algorithms, Assembly, Annealing, Cryptography, and Experimental Computing~\cite{qosf_project_list}.
Each repository is uniquely identified using the format \texttt{<Owner/Repository>}.

For each repository, we collected repository-level attributes such as name, primary programming language, license, repository size (KB), and total number of reported bugs.
At the issue level, we extracted metadata including issue identifiers, titles, descriptions, labels, timestamps, and comments, capturing both technical and maintenance-related characteristics.

Repositories with no recorded bugs were retained to ensure dataset completeness, as their documentation, testing artifacts, and development activity contribute to assessing software maturity and quality practices.

\begin{table}[h!]
\centering
\caption{Summary of analyzed repositories by category, including repository count, total bugs, and mean size.}
\label{tab:repo_summary}
\begin{tabular}{lccc}
\hline
\textbf{Category} & \textbf{Repositories} & \textbf{Total Bugs} & \textbf{Mean Size (KB)} \\
\hline
Quantum Full Stack Libraries & 19 & 10,642 & 39,871 \\
Quantum Simulators & 43 & 3,804 & 29,376 \\
Quantum Annealing & 17 & 1,328 & 7,314 \\
Quantum Algorithms & 19 & 3,273 & 23,947 \\
Quantum Compilers & 10 & 6,551 & 32,917 \\
Quantum Assembly & 4 & 292 & 4,724 \\
Quantum Cryptography & 4 & 1,221 & 81,604 \\
Quantum Experimental Computing & 8 & 2,239 & 9,047 \\
\hline
\end{tabular}
\end{table}

\subsection{Bug Classification Dimensions}
\label{bugclassificationdim}
Bug reports were classified using a multi-dimensional rule-based framework that systematically labels each issue across five key dimensions: \textit{bug type}, \textit{category}, \textit{severity}, \textit{quality attribute}, and \textit{quantum-specific subtype}.  
The framework integrates domain-specific keyword dictionaries, rule-based heuristics, and TF–IDF–based contextual scoring to distinguish between classical and quantum-domain defects, while capturing fine-grained quality and severity characteristics.  
Quantum-related bugs are further subdivided into circuit-, gate-, measurement-, noise-, transpilation-, and execution-level categories to reflect NISQ-specific challenges.

Importantly, the framework generalizes across heterogeneous ecosystems, incorporating rule sets that recognize terminology and labeling patterns from frameworks such as \textit{Qiskit}, \textit{Cirq}, \textit{PyQuil}, and \textit{D-Wave Ocean}.  
The classification approach follows the framework evaluated in~\cite{yousuf2026bug}, which demonstrated accuracies of 0.81–0.85 and macro–F1 values of 0.66–0.77.
Using an established framework enables continuity with earlier studies and supports transparent, comparable empirical analysis

\subsection{Static Code and Documentation Analysis}\label{staticcodeanddocanalysis}

To assess maintainability and structural quality, we performed static code analysis on 60 Python-based repositories.  
Cyclomatic complexity ($M$) was computed using McCabe’s formula:

\begin{equation}
M = E - N + 2P
\end{equation}

where $E$ represents control-flow edges, $N$ represents nodes, and $P$ denotes the number of connected components.  
The average repository complexity was computed as:

\begin{equation}
\bar{C} = \frac{C_{\text{total}}}{f}
\end{equation}

where $C_{\text{total}}$ is the sum of per-function complexities, and $f$ is the number of syntactically valid files.  
Code quality was measured using \texttt{Pylint} scores, which range from $-10.0$ (poor quality) to $10.0$ (excellent)~\cite{pylint2025}.
Aggregated complexity and code quality statistics across quantum software domains are summarized in Table~\ref{tab:static_summary}.

The full processing workflow is formalized in Algorithm~\ref{alg:static_analysis}, which encodes the repository-level static analysis pipeline.
The algorithm identifies Python source files, filters syntactically invalid files, and extracts function- and method-level blocks using \texttt{radon}.
For each block, McCabe cyclomatic complexity is computed and subsequently aggregated at the repository level along with block counts.
An overall code quality score is then obtained by executing \texttt{pylint} on each repository.

In addition to static code metrics, the pipeline derives documentation indicators—including README quality, changelog presence, docstring coverage, and the existence of \texttt{docs/} and \texttt{examples/} directories—as well as testing indicators such as the presence of test files and coverage configuration.
All repository-level measurements are persisted to support downstream aggregation and statistical modeling.

\begin{table}[h!]
\centering
\caption{Summary of static analysis results for Python-based repositories (mean ± SD).}
\label{tab:static_summary}
\begin{tabular}{lccc}
\hline
\textbf{Category} & \textbf{Pylint Score} & \textbf{Avg. Cyclomatic Complexity} & \textbf{Blocks} \\
\hline
Quantum Full Stack Libraries & 6.83 ± 2.3 & 2.81 ± 1.2 & 1,832 ± 965 \\
Quantum Simulators & 5.21 ± 3.1 & 2.64 ± 1.3 & 1,245 ± 812 \\
Quantum Algorithms & 7.34 ± 1.9 & 3.11 ± 1.4 & 1,007 ± 578 \\
Quantum Compilers & 6.95 ± 2.5 & 3.02 ± 1.1 & 986 ± 433 \\
Quantum Annealing & 6.47 ± 2.0 & 2.56 ± 1.0 & 591 ± 270 \\
Quantum Experimental Computing & 4.23 ± 2.8 & 2.78 ± 0.9 & 863 ± 541 \\
\hline
\end{tabular}
\end{table}

\subsubsection{Documentation Quality Evaluation}\label{docquality}

To assess the completeness and usability of documentation across quantum software repositories, we implemented an automated scoring framework that evaluates five core documentation components.  
Each component was quantified using structured scoring functions to ensure objectivity and comparability across projects.

The following metrics were computed for each repository:

\begin{itemize}
    \item \textbf{README Completeness} ($S_{\text{README}}$): evaluated on a scale of 1–5 based on the presence of essential sections such as project overview, installation instructions, usage examples, and badges:
    \begin{equation}
      S_{\text{README}} \in [1, 5]
    \end{equation}

    \item \textbf{CHANGELOG Presence} ($S_{\text{CH}}$): scored as 5 if a \texttt{CHANGELOG.md} file was present and 1 otherwise:
    \begin{equation}
    S_{\text{CH}} =
    \begin{cases}
    5, & \text{if \texttt{CHANGELOG.md} exists} \\
    1, & \text{otherwise}
    \end{cases}
    \end{equation}

    \item \textbf{Inline Documentation} ($S_{\text{DOC}}$): evaluated by checking docstrings in Python files.  
    Let $N$ be the total number of Python files and $D$ the number with docstrings:
    \begin{equation}
    S_{\text{DOC}} =
    \begin{cases}
    5, & \text{if } \frac{D}{N} > 0.5 \\
    3, & \text{otherwise}
    \end{cases}
    \end{equation}

    \item \textbf{User Documentation} ($S_{\text{USR}}$): scored based on the existence of a \texttt{docs/} directory:
    \begin{equation}
    S_{\text{USR}} =
    \begin{cases}
    5, & \text{if \texttt{docs/} directory exists} \\
    1, & \text{otherwise}
    \end{cases}
    \end{equation}

    \item \textbf{Tutorials/Examples} ($S_{\text{TUT}}$): assessed via the presence of an \texttt{examples/} directory:
    \begin{equation}
    S_{\text{TUT}} =
    \begin{cases}
    5, & \text{if \texttt{examples/} directory exists} \\
    1, & \text{otherwise}
    \end{cases}
    \end{equation}
\end{itemize}

Binary indicators were also computed to capture \textbf{Testing Presence} ($T$) and \textbf{Coverage Configuration} ($C$):
\[
T, C \in \{0, 1\}
\]
where 1 indicates presence and 0 absence.  

Each metric thus provides a quantitative snapshot of the documentation and testing ecosystem across repositories, enabling cross-project comparison of maintainability and usability indicators.
 
Table~\ref{tab:doc_test_summary} reports the aggregated summary for all Python-based repositories, presenting the mean scores, standard deviations, and presence percentages for each metric.  
These statistics highlight the relative maturity of documentation and testing practices across the quantum software landscape.

\begin{table}[h]
\centering
\caption{Documentation and testing metrics for Python repositories.}
\label{tab:doc_test_summary}
\begin{tabular}{lccc}
\hline
\textbf{Metric} & \textbf{Mean Score} & \textbf{Std. Dev.} & \textbf{Presence (\%)} \\
\hline
README Completeness ($S_{\text{README}}$) & 26.4 & 13.2 & 92.7 \\
CHANGELOG Presence ($S_{\text{CH}}$) & 1.9 & 0.6 & 54.1 \\
Inline Docstrings ($S_{\text{DOC}}$) & 4.1 & 0.9 & 68.3 \\
User Documentation ($S_{\text{USR}}$) & 3.8 & 1.2 & 63.5 \\
Tutorials/Examples ($S_{\text{TUT}}$) & 2.9 & 1.5 & 47.2 \\
Testing Presence ($T$) & – & – & 61.0 \\
Coverage Tools ($C$) & – & – & 27.8 \\
\hline
\end{tabular}
\end{table}

Documentation and testing metrics were computed at the repository level using a structured scoring model that captures both quantitative scores and binary indicators.
This model provides a consistent and interpretable measure of documentation maturity and testing integration across the quantum software ecosystem, supporting comparative analysis across repositories and domains.

\begin{algorithm}[h!]
\caption{Static Code and Documentation Analysis}
\label{alg:static_analysis}
\begin{algorithmic}[1]
\Require List of repositories $R = \{r_1, r_2, \dots, r_n\}$
\Ensure Per-repository metrics and aggregated statistics for complexity, quality, and documentation

\For{each repository $r \in R$}
    \State $F \gets$ \textsc{FindPythonFiles}($r$)
    \State $F_{\text{valid}} \gets$ \textsc{FilterValidPythonFiles}($F$) \Comment{Skip syntactically invalid files}
    \State $C_{\text{total}} \gets 0$, $f \gets |F_{\text{valid}}|$, $B \gets 0$
    
    \For{each file $f_i \in F_{\text{valid}}$}
        \State $blocks \gets$ \textsc{RadonGetBlocks}($f_i$)
        \For{each block $b \in blocks$}
            \State $M_b \gets$ \textsc{ComputeMcCabe}($b$)
            \State $C_{\text{total}} \gets C_{\text{total}} + M_b$
            \State $B \gets B + 1$
        \EndFor
    \EndFor
    \State $\bar{C} \gets \frac{C_{\text{total}}}{\max(1, f)}$ \Comment{Average cyclomatic complexity}
    \State $S_{\text{Pylint}} \gets$ \textsc{RunPylint}($r$) \Comment{Static code quality $[-10,10]$}
    
    \State $T \gets 1$ if \textsc{ContainsTestFiles}($r$) else $0$
    \State $C \gets 1$ if \textsc{CoverageConfigured}($r$) else $0$
    
    \Comment{Compute documentation scores}
    \State $S_{\text{README}} \gets$ \textsc{ScoreREADME}($r$)
    \State $S_{\text{CH}} \gets 5$ if \texttt{CHANGELOG.md} exists else $1$
    \State $(N, D) \gets$ \textsc{CountFilesWithDocstrings}($F_{\text{valid}}$)
    \State $S_{\text{DOC}} \gets 5$ if $\frac{D}{N} > 0.5$ else $3$
    \State $S_{\text{USR}} \gets 5$ if \texttt{docs/} directory exists else $1$
    \State $S_{\text{TUT}} \gets 5$ if \texttt{examples/} or \texttt{tutorials/} directory exists else $1$
    
    \State Save per-repository results:
    \[
    \{ S_{\text{Pylint}}, \bar{C}, B, T, C, S_{\text{README}}, S_{\text{CH}}, S_{\text{DOC}}, S_{\text{USR}}, S_{\text{TUT}} \}
    \]
\EndFor

\end{algorithmic}
\end{algorithm}

\subsection{Validation and Reliability of the Classification Framework}\label{validation}

The present study extends the rule-based classification methodology introduced in~\cite{yousuf2026bug} by applying it to a substantially larger, multi-ecosystem dataset and augmenting it with additional verification steps. 
The original framework integrates weighted keyword heuristics, TF--IDF similarity, and domain-specific rules to categorize quantum software issues across five analytical dimensions: \textit{bug type}, \textit{bug category}, \textit{severity}, \textit{quality attribute}, and \textit{quantum-specific subtype}. 
Its prior evaluation on 12{,}910 issues from 36 Qiskit repositories, together with a stratified subset of 4{,}984 manually annotated issues, reported accuracies of 0.81--0.85 and macro--F1 scores of 0.66--0.77, outperforming several machine-learning baselines including Logistic Regression, Random Forest, and Gradient Boosting. 
Inter-annotator agreement among three reviewers yielded Cohen's~$\kappa$ values between 0.78 and 0.82, indicating substantial to near-perfect agreement~\cite{landis1977measurement}.

To assess cross-ecosystem behavior, the framework was also applied to 4{,}613 issues from 11 repositories across the \textit{Cirq}, \textit{PyQuil}, and related ecosystems. 
At the level of high-level bug types, no statistically significant deviation was observed across ecosystems (\(\chi^2 = 4.75, p = 0.093, V = 0.019\)), with broadly similar proportions of classical (\(\sim 67\%\)) and quantum-specific (\(\sim 27\text{--}30\%\)) defects. 
At finer levels of analysis, ecosystem-specific variation appeared for \textit{Bug Category} (\(\chi^2 = 277.84, p < 0.0001, V = 0.126\)), 
\textit{Severity} (\(\chi^2 = 151.95, p < 0.0001, V = 0.093\)), 
\textit{Quality Attribute} (\(\chi^2 = 214.39, p < 0.0001, V = 0.111\)), and 
\textit{Quantum-Specific Subtype} (\(\chi^2 = 94.33, p < 0.0001, V = 0.227\)), reflecting known differences in abstraction layers and tooling across frameworks. 
Overall, these results suggest that the methodology is robust enough for multi-ecosystem empirical analyses while remaining sensitive to domain-specific defect patterns.

In extending the methodology to a corpus of 123 repositories and more than 32{,}000 issues, this study conducted three supplementary verification procedures to evaluate its behavior at scale:

\begin{enumerate}[label=(\alph*)]
    \item \textbf{Manual Sampling Validation:} A stratified random sample of classified issues was manually reviewed to assess the consistency of label assignments across analytical dimensions.

    \item \textbf{Cross-Domain Consistency Check:} Distributions of defect types and categories were compared across repository domains to examine whether coherent, interpretable patterns emerged in different project contexts.

    \item \textbf{Deterministic Reproducibility:} The classification pipeline was executed multiple times on identical inputs to confirm that it produced stable, deterministic outputs.
\end{enumerate}

These supplementary checks, while not intended to replace full-scale re-annotation, provide supporting evidence that the framework operates consistently on the expanded multi-ecosystem corpus and remains suitable for the downstream analyses of defect characteristics, testing practices, and software quality trends presented in this work.

\subsection{Data Aggregation and Statistical Analysis}\label{aggregation_analysis}

Following bug classification and static analysis, all repository-level and issue-level data were aggregated to enable statistical evaluation of trends and relationships across the quantum software ecosystem.  
The analyses combined descriptive, correlational, and regression-based techniques to identify patterns and quality drivers in quantum software development.  
All statistical analyses were implemented in Python using \texttt{NumPy}, \texttt{Pandas}, \texttt{SciPy}, and \texttt{StatsModels} to ensure transparency and reproducibility.

\subsubsection*{Data Aggregation and Normalization}
All extracted metrics were normalized at the repository level to ensure comparability across projects of varying sizes and activity levels.  
For each repository $r$, the normalized bug density was computed as:
\begin{equation}
B_r = \frac{\text{Number of Bugs}_r}{\text{Total Files}_r}
\end{equation}
and the testing ratio as:
\begin{equation}
T_r = \frac{\text{Test Files}_r}{\text{Total Python Files}_r}
\end{equation}
These normalized indicators were used to measure code quality and testing intensity independent of project scale.

\subsubsection*{Descriptive and Comparative Analysis:}
We computed descriptive statistics (mean, median, and standard deviation) for all major quantitative variables, including bug counts, cyclomatic complexity, Pylint scores, and documentation quality.  
To examine inter-category variations, we applied pairwise comparisons using non-parametric tests (Kruskal–Wallis and Mann–Whitney U tests) for variables not normally distributed, verified via the Shapiro–Wilk test ($p<0.05$).  
This allowed robust comparisons of software quality indicators across repository categories without assuming Gaussian data distribution.

\subsubsection*{Correlation and Regression Analysis:}
To quantify relationships among metrics, we first employed Spearman’s rank correlation ($\rho$) to examine associations between bug density, code complexity, testing, and documentation quality.  
All variables exhibiting significant correlation ($p<0.05$) were subsequently included in regression modeling to estimate effect magnitudes and directions.

Because the dependent variable (\textit{number of reported bugs}) is a non-negative count exhibiting over-dispersion (variance exceeding the mean), we used a Negative-Binomial Generalized Linear Model (GLM) with a log link to ensure distributional robustness.  
The final regression model was specified as:
\begin{equation}
\begin{aligned}
\text{Number\_of\_Bugs} \sim\;& Q(\textit{'has\_tests'}) 
+ Q(\textit{'Average\_Pylint\_Score'}) \\
&+ Q(\textit{'Cyclomatic\_Complexity'}) 
+ Q(\textit{'Number\_of\_Blocks'})
\end{aligned}
\end{equation}
where:
\begin{itemize}
    \item $Q(\textit{'\_has\_tests'})$ is a binary indicator of testing presence (1 = tests available),
    \item $Q(\textit{'Average\_Pylint\_Score'})$ captures average static-analysis quality,
    \item $Q(\textit{'Cyclomatic\_Complexity'})$ represents structural code complexity, and
    \item $Q(\textit{'Number\_of\_Blocks'})$ proxies repository size.
\end{itemize}

The model was fitted across 60 Python-based repositories for which all predictors were available.  
Model diagnostics confirmed adequate fit (deviance = 174.26, Pearson $\chi^2$ = 100, $p < 0.001$), with low multicollinearity (variance inflation factors(VIF) $<2.0$ for all predictors) and pseudo $R^2_{\text{CS}} = 0.7323$, indicating strong explanatory power.  
Residual analysis revealed no heteroskedasticity or influential outliers.

\subsubsection*{Temporal and Distributional Analysis:}
Temporal bug trends were analyzed using year-based aggregation to capture the evolution of bug occurrences, severity, and types over time.  
We employed moving averages and exponential smoothing to reduce noise in year-over-year bug counts and to detect long-term ecosystem trends.  
Distributional plots were generated to visualize variation across repository categories.

All resulting correlations, regression coefficients, and diagnostics are presented in Section~\ref{sec:resultsandobservations}, 
including the complete negative-binomial regression summary in Table~\ref{tab:nb_regression}.

\section{Results and Observations}
\label{sec:resultsandobservations}

This section presents the empirical findings derived from the large-scale analysis of 123 open-source quantum software repositories, encompassing 32,296 classified issues.  
The results integrate outputs from the validated rule-based classification framework~\cite{yousuf2026bug}, static code and documentation metrics, and repository-level metadata to address the twelve RQs defined in Section~\ref{intro}.  
Each analysis connects bug characteristics, testing practices, and documentation quality to core software quality attributes such as reliability, maintainability, and performance.

To ensure coherence, the findings are organized into seven thematic dimensions corresponding to the RQs:

\begin{itemize}
    \item \textbf{RQ1–RQ2:} Bug evolution and distribution ,  analyzing temporal patterns and ecosystem-level defect trends.
    \item \textbf{RQ3–RQ5:} Quality attributes and reliability ,  examining how defects map to software quality dimensions and how these differ between classical and quantum-specific issues.
    \item \textbf{RQ6–RQ7:} Programming languages and documentation ,  investigating the influence of language ecosystems and documentation completeness on bug occurrence and resolution.
    \item \textbf{RQ8–RQ10:} Testing, complexity, and bug tracking ,  evaluating relationships between testing practices, code complexity, and defect density.
    \item \textbf{RQ11–RQ12:} Quantum-specific bug distribution ,  characterizing the types and technical impact of quantum-domain defects.
\end{itemize}

Each subsection reports quantitative results, followed by interpretive synthesis and key observations.  
Together, these analyses provide a data-driven characterization of defect patterns and quality drivers across the quantum software ecosystem, highlighting the gradual engineering maturity of QSE.

\subsection{Bug Evolution and Distribution in Quantum Software Repositories}
\label{subsec:bug_evolution_distribution}

Understanding how bugs evolve over time and across repository types provides insight into the maturity of the quantum software ecosystem.  
As the field transitions from experimental prototypes to production-ready systems, repositories increasingly reflect hybrid challenges spanning both classical and quantum layers.  
Tracking longitudinal bug patterns thus reveals how testing frameworks, engineering practices, and community participation influence software reliability and maintainability.

To analyze these dynamics, we examined bug trends across 123 repositories spanning a twelve-year period (2012–2024).  
The following research question guided this analysis.

\textbf{RQ1:} \textit{How have bug counts and categories in open-source quantum software evolved over time, and how do these trends differ across repository types?}

\textit{Subquestions:}
\begin{enumerate}
  \item \textbf{RQ-1.1:} Which repository categories (e.g., full-stack libraries, compilers, simulators) contributed most to changes in bug frequency over time?
  \item \textbf{RQ-1.2:} Which bug types (compatibility, functional, quantum-specific, etc.) are most prevalent, and how has their prevalence changed from 2012 to 2024?
  \item \textbf{RQ-1.3:} What do these temporal and categorical patterns suggest about the maturity and reliability of the quantum software ecosystem?
\end{enumerate}

Across all repositories, 73\% of issues were classified as classical and 23\% as quantum-specific, reaffirming the hybrid nature of current quantum software.  
This distribution reflects that while quantum computation introduces domain-specific reliability challenges, classical infrastructure continues to dominate overall development complexity.

\subsubsection*{(1) Temporal Evolution of Bugs:}
\begin{figure}[h!]
    \centering
    \includegraphics[width=0.9\linewidth]{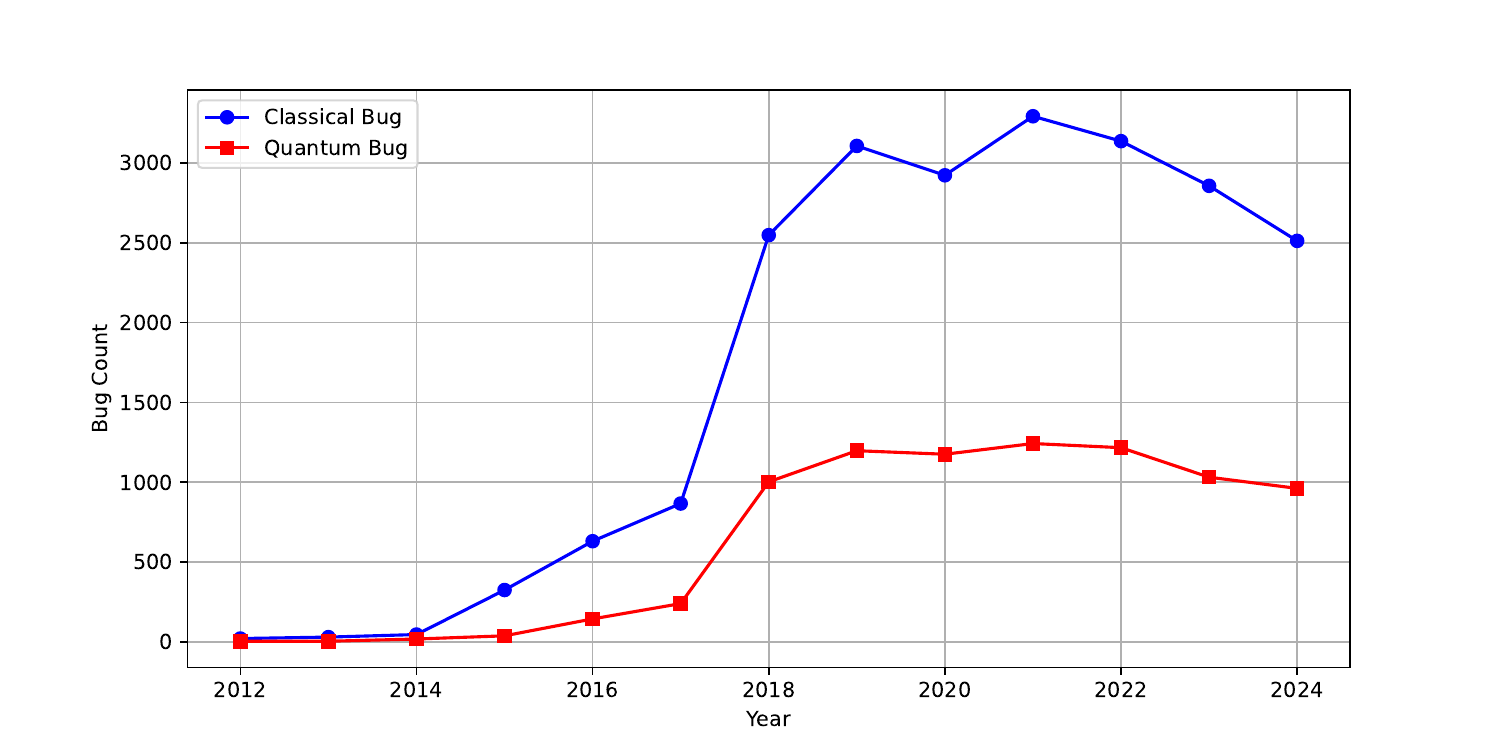}
    \caption{Year-wise distribution of classical and quantum bugs across all repositories (2012--2024).}
    \label{fig:trendrepos}
\end{figure}

Figure~\ref{fig:trendrepos} presents the annualized distribution of classical and quantum bug reports derived from 123 repositories, showing empirically observed trends from 2012–2024. In the early period (2012--2016), reported defects were sparse (classical: 20~→~630; quantum: 3~→~143), mirroring the infancy of quantum programming environments.  
Between 2017 and 2021, bug counts increased sharply as large frameworks such as \textit{Qiskit}, \textit{Cirq}, and \textit{PennyLane} emerged, expanding participation and exposing more defects through systematic testing.  
During this growth phase, classical bugs peaked at 3,293 and quantum-specific at 1,242 (2021).  
After 2021, both trends stabilized (classical: 2,512; quantum: 961 by 2024), signaling improved testing pipelines, community-driven maintenance, and greater engineering maturity.

\subsubsection*{(2) Bug Trends Across Repository Categories:}
\begin{figure}[h!]
    \centering
    \includegraphics[width=0.99\linewidth]{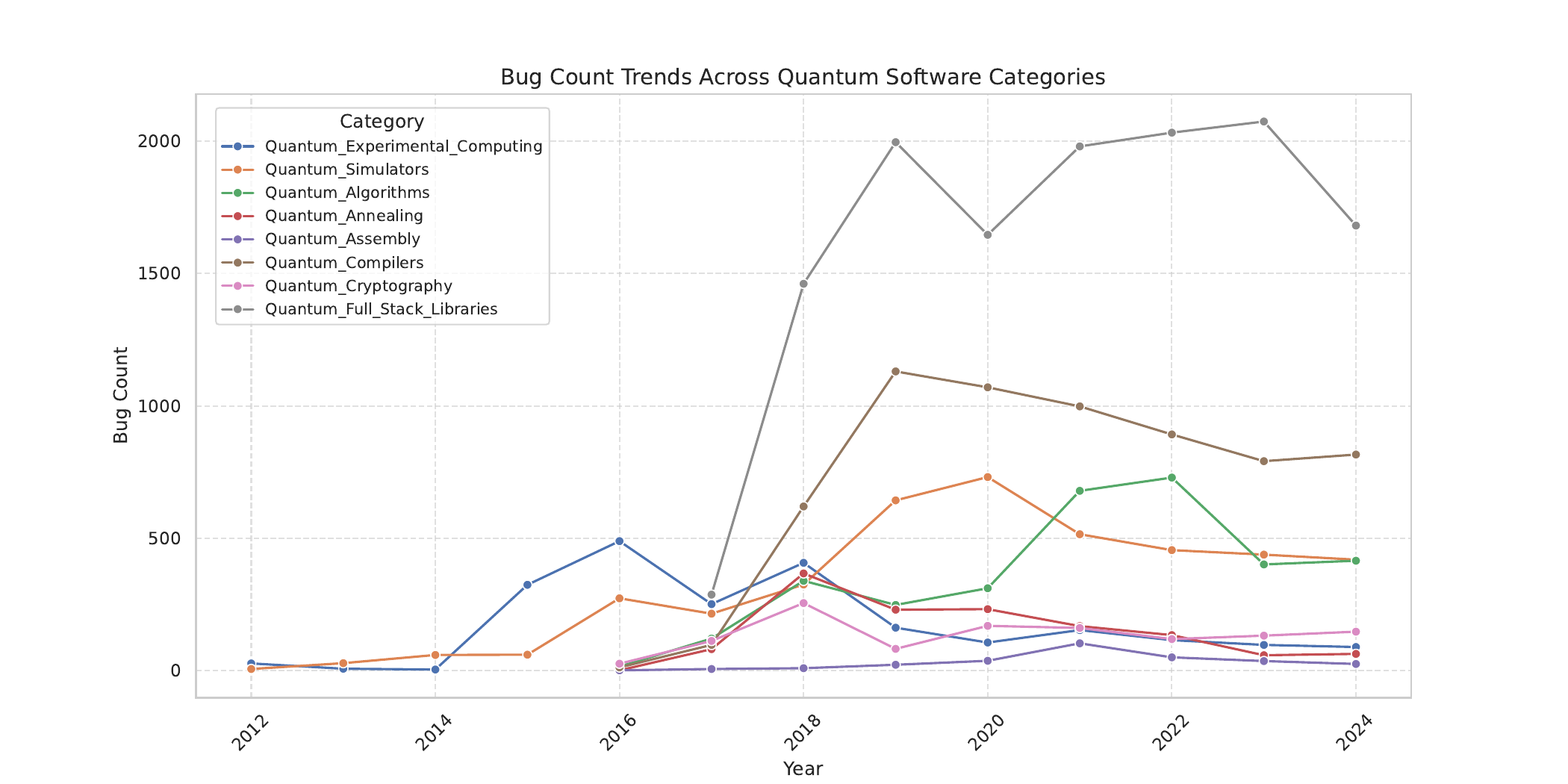}
    \caption{Year-wise bug trends across major quantum software repository categories (2012--2024).}
    \label{fig:trendcategories}
\end{figure}

Distinct repository categories exhibit different longitudinal trajectories (Figure~\ref{fig:trendcategories}).  
In the formative phase (2012--2016), \textit{Quantum Simulators} and \textit{Experimental Computing} tools recorded the highest bug counts, reflecting foundational challenges in simulation accuracy and early hardware interfacing.  
From 2017 to 2021, bug activity surged across nearly all categories, particularly in \textit{Full-Stack Libraries} and \textit{Quantum Compilers}.  
Classical bugs in Full-Stack Libraries rose from 215 (2017) to 1,453 (2023), while quantum-specific bugs increased from 60 to 557.  
This expansion coincides with the integration of complex toolchains and hybrid execution backends.  
Post-2021, most categories show stabilizing or declining trends, suggesting increased modularization and more mature testing frameworks.  
Nevertheless, \textit{Quantum Compilers} and \textit{Simulators} remain high-defect areas, underscoring persistent engineering difficulties in gate-level optimization, backend translation, and noise modeling.

\subsubsection*{(3) Prevalent Bug Categories and Distribution Patterns:}
\begin{figure}[h!]
    \centering
    \includegraphics[width=0.99\linewidth]{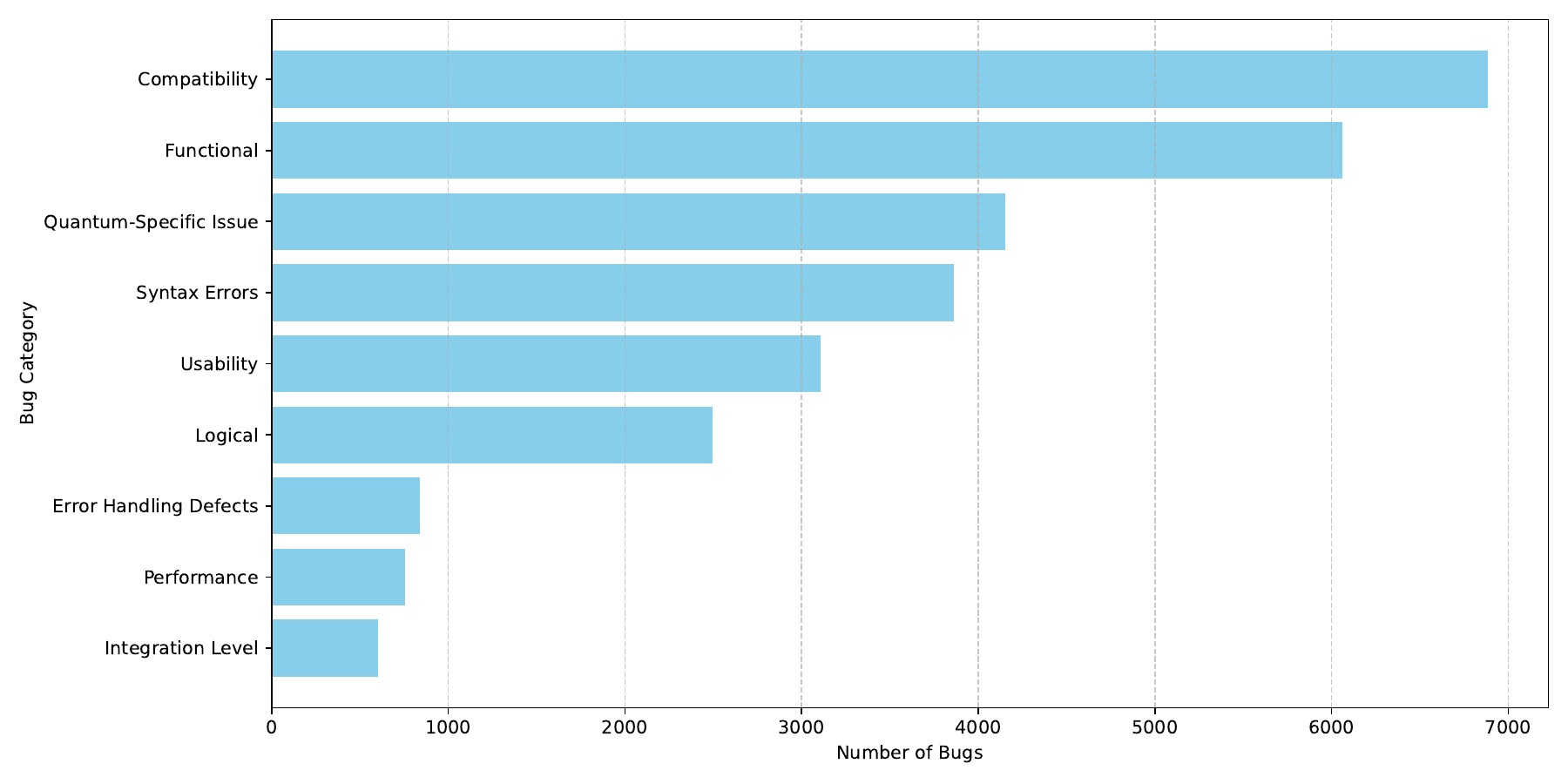}
    \caption{Distribution of bug categories across analyzed quantum software repositories.}
    \label{fig:bug_categories}
\end{figure}

The distribution of bug types reveals dominant reliability challenges across repositories.  
As shown in Figure~\ref{fig:bug_categories}, \textbf{compatibility} (6,883), \textbf{functional} (6,063), and \textbf{quantum-specific} (4,152) defects are most prevalent, highlighting persistent issues of integration, correctness, and domain-specific reliability.  
Compatibility bugs stem primarily from mismatched classical–quantum interfaces, while functional bugs relate to algorithmic correctness.  
Quantum-specific issues, such as gate-mapping faults and qubit mismanagement, require domain-aware debugging and specialized tooling.  
Syntax (3,863) and usability (3,106) defects remain frequent but show gradual reduction, reflecting progress in developer tooling and documentation.

Table~\ref{tab:bugs_distribution} details the distribution of defect categories by repository type.  
\textbf{Full-Stack Libraries} exhibit the highest counts across nearly all categories (compatibility:~3,253; functional:~2,301; quantum-specific:~1,974), followed by \textbf{Quantum Compilers} (compatibility:~1,527; functional:~1,183; quantum-specific:~1,141).  
Integration-heavy frameworks are thus the most error-prone due to complex orchestration between quantum and classical components.  
\textbf{Quantum Simulators} and \textbf{Experimental Computing} tools show notable performance and logic issues, whereas \textbf{Quantum Assembly} repositories contain the fewest defects, reflecting their narrower functional scope.

\begin{table*}[t]
\centering
\caption{Bug distribution across quantum software categories (BE: Boundary Errors, CE: Calculation Errors, CommE: Communication Errors, Comp: Compatibility, CD: Critical Bugs, EH: Error Handling, Func: Functional, IL: Integration Level, LE: Logical Errors, OOB: Out of Bound, Perf: Performance, QSI: Quantum-Specific Issues, RE: Runtime Errors, Sec: Security, SE: Syntax Errors, Usab: Usability).}
\label{tab:bugs_distribution}
\renewcommand{\arraystretch}{1.05}
\setlength{\tabcolsep}{3pt}
\scriptsize
\begin{tabular}{lcccccccccccccccc}
\hline
\textbf{Category} & \textbf{BE} & \textbf{CE} & \textbf{CommE} & \textbf{Comp} & \textbf{CD} & \textbf{EH} & \textbf{Func} & \textbf{IL} & \textbf{LE} & \textbf{OOB} & \textbf{Perf} & \textbf{QSI} & \textbf{RE} & \textbf{Sec} & \textbf{SE} & \textbf{Usab} \\
\hline
Q Al & 22 & 24 & 8 & 723 & 47 & 90 & 936 & 36 & 105 & 23 & 73 & 298 & 26 & 8 & 353 & 348 \\
Q An & 29 & 7 & 8 & 122 & 25 & 41 & 242 & 42 & 141 & 12 & 39 & 24 & 28 & 16 & 134 & 159 \\
Q As & 3 & 4 & 31 & 4 & 4 & 4 & 55 & 4 & 20 & 11 & 5 & 51 & 8 & -- & 40 & 27 \\
Q Comp & 30 & 46 & 20 & 1527 & 82 & 174 & 1183 & 89 & 249 & 33 & 166 & 1141 & 40 & 42 & 677 & 611 \\
Q Cryp & 4 & 17 & 16 & 187 & 4 & 21 & 176 & 57 & 115 & 5 & 41 & 18 & 17 & 19 & 189 & 169 \\
Q EC & 25 & 36 & 28 & 297 & 58 & 83 & 367 & 73 & 253 & 14 & 74 & 23 & 88 & 13 & 298 & 233 \\
Q FSL & 61 & 98 & 46 & 3253 & 208 & 295 & 2301 & 224 & 560 & 72 & 252 & 1974 & 102 & 84 & 1618 & 1252 \\
Q Sim & 21 & 42 & 29 & 743 & 56 & 133 & 803 & 76 & 248 & 12 & 105 & 623 & 39 & 25 & 554 & 307 \\
\hline
\end{tabular}
\end{table*}

\textbf{Key Observations.}
\begin{itemize}
    \item \textbf{Compatibility issues} dominate Full-Stack (3,253), Compiler (1,527), and Simulator (743) repositories, reflecting persistent integration and dependency challenges.
    \item \textbf{Quantum-specific defects} are concentrated in Full-Stack (1,974) and Compiler (1,141) projects, highlighting difficulties in circuit synthesis and hardware adaptation.
    \item \textbf{Functional and performance bugs} remain frequent in Full-Stack (2,301) and Simulator (803) repositories, underscoring algorithmic and efficiency bottlenecks.
    \item \textbf{Usability and documentation gaps} persist in Full-Stack (1,252) and Compiler (611) projects, indicating the need for enhanced developer support and structured documentation.
    \item \textbf{Security defects}, though infrequent, are non-negligible (84 in Full-Stack, 42 in Compilers), suggesting emerging risks in hybrid architectures.
\end{itemize}

\subsubsection*{(4) Classical vs. Quantum-Specific Bugs:}
\begin{figure}[h!]
    \centering
    \includegraphics[width=0.99\linewidth]{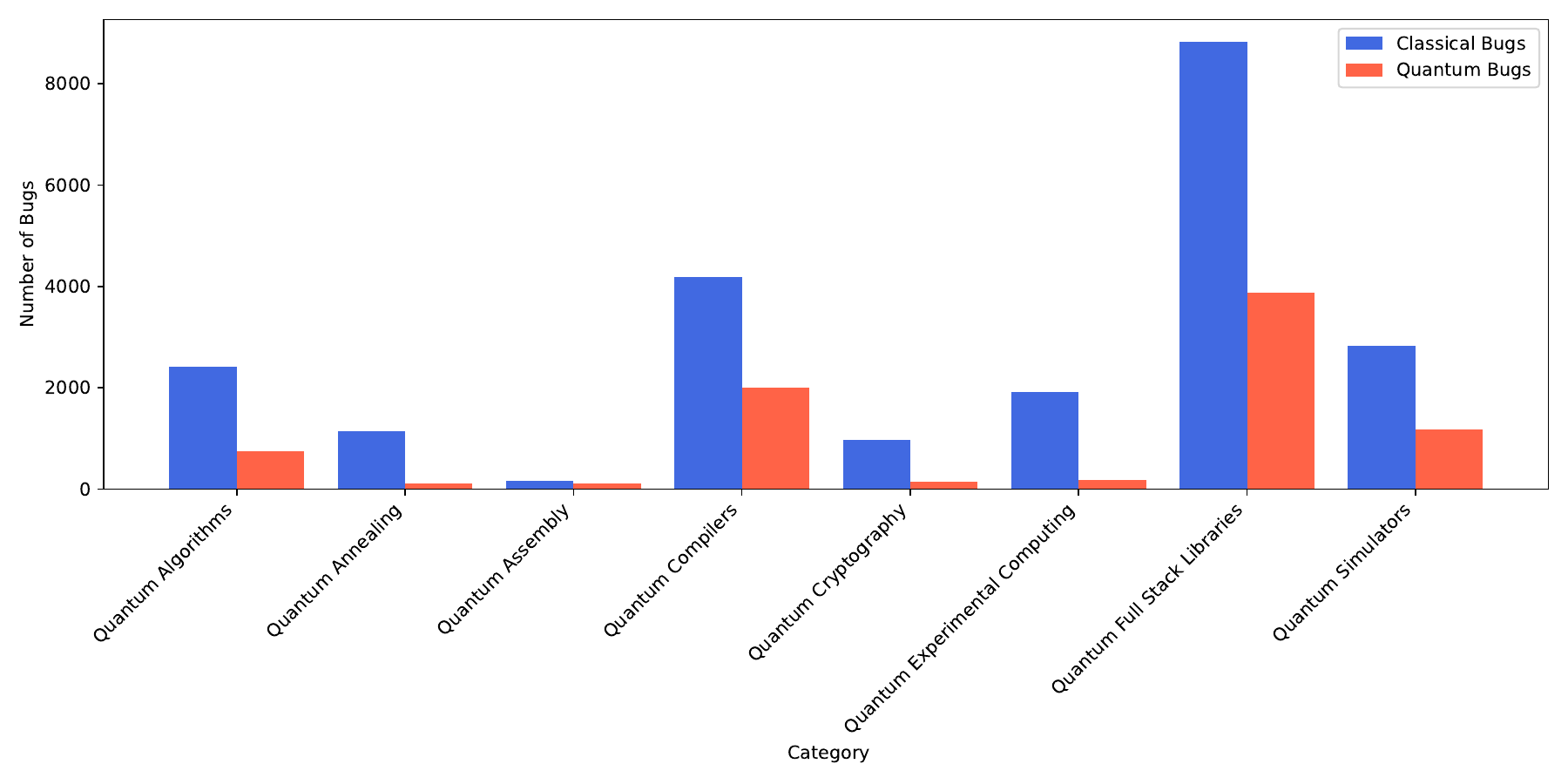}
    \caption{Distribution of classical and quantum-specific bugs across repository categories.}
    \label{fig:bugstype_dist_category}
\end{figure}

Figure~\ref{fig:bugstype_dist_category} visualizes classical versus quantum-specific bug counts across repository categories.  
\textit{Full-Stack Libraries} contain the largest share of both classical (8,822) and quantum-specific (3,866) bugs, followed by \textit{Quantum Compilers} (2,005 quantum-specific).  
This reinforces that large, integrated systems face dual reliability challenges, classical interoperability and quantum correctness.

\subsubsection*{(5) Interpretation and Implications:}
The combined temporal and categorical trends indicate a gradual maturation of QSE practices.  
The post-2021 decline in reported bugs reflects improved testing frameworks, standardized development pipelines, and stronger community quality control.  
However, high defect densities in compilers and full-stack frameworks underscore persistent integration challenges and technical debt.  
Key implications include:
\begin{itemize}
    \item \textbf{Tooling and verification:} develop hybrid debugging and validation tools for compilers and transpilers.
    \item \textbf{Standardization:} adopt interface and compatibility standards to mitigate integration-level failures.
    \item \textbf{Developer support:} improve documentation, tutorials, and static-analysis aids to reduce usability and syntax errors.
    \item \textbf{Specialized testing:} integrate quantum-specific test suites for gate-level and hardware-coupled defects.
\end{itemize}




\begin{rqanswer}{1}
\textbf{Answer to RQ-1.1:} The longitudinal analysis shows a clear evolution of quantum software repositories from small experimental prototypes to structured, multi-layered frameworks with regular testing and issue tracking practices. 
\textit{Full-stack libraries} and \textit{compilers} account for the largest increases in defect volume, particularly during 2017--2021 when major SDKs matured and hardware--software integration intensified. 
After 2021, defect growth stabilized, indicating increasing process maturity, improved CI adoption, and more stable release cycles across the ecosystem.

\textbf{Answer to RQ-1.2:} Compatibility, functional, and quantum-specific issues constitute more than 65\% of all reported defects, reflecting persistent challenges in coordinating classical and quantum components. 
In contrast, syntax- and usability-related defects show a consistent decline over time, suggesting improvements in developer tooling, documentation quality, and automated testing infrastructures that mitigate early-stage programming errors.

\textbf{Answer to RQ-1.3:} Overall, the temporal and categorical patterns point to a maturing QSE ecosystem characterized by more disciplined testing, stronger community participation, and emerging standardization practices. 
At the same time, recurring issues in compilers, transpilation, and cross-stack integration highlight ongoing needs for specialized debugging methods, quantum-aware testing, and more robust hybrid verification pipelines tailored to NISQ-era constraints.

Taken together, these findings indicate that while reliability, tooling, and development workflows have steadily improved, substantial challenges remain in compiler optimization, hardware calibration, and multi-layer defect detection, factors that continue to shape the trajectory of quantum software engineering.
\end{rqanswer}

\subsection{Bug Severity and Debugging}
\label{subsec:bug_severity_debugging}

Bug severity shapes both the difficulty of debugging and the operational risk of quantum software. This subsection
quantifies severity distributions across repository categories and draws implications for targeted debugging,
verification, and maintenance strategies. It addresses the following research question:

\textbf{RQ2:} \textit{How does bug severity vary across different categories of quantum software repositories, and what implications does this have for debugging and reliability in quantum computing?}

\begin{figure}[h!]
    \centering
    \includegraphics[width=0.99\linewidth]{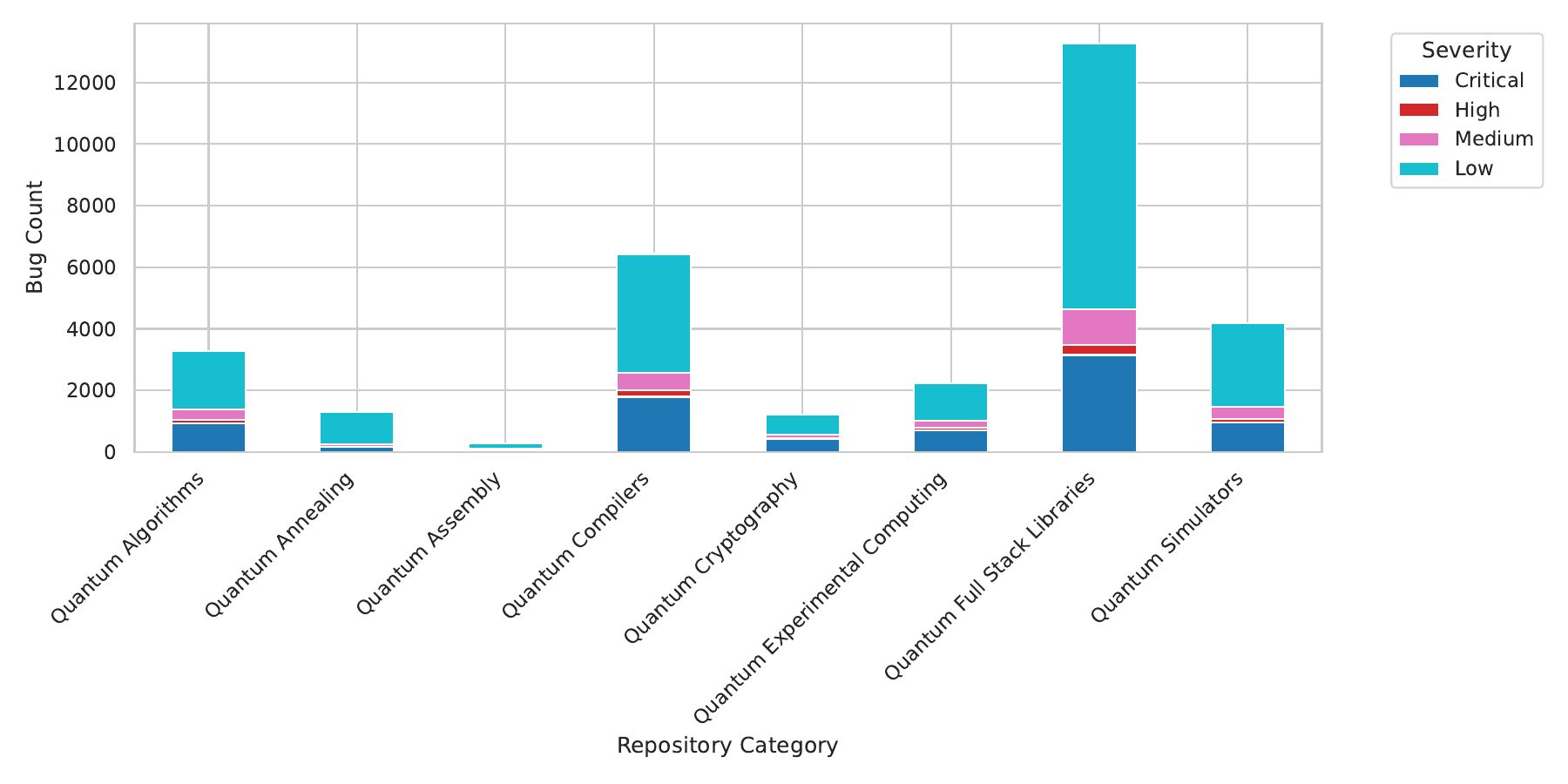}
    \caption{Distribution of bug severity levels (Critical, High, Medium, Low) across eight categories of quantum software repositories.}
    \label{fig:distbugcat}
\end{figure}

Figure~\ref{fig:distbugcat} shows the proportional distribution of four severity levels (\textit{Low}, \textit{Medium},
\textit{High}, \textit{Critical}) across eight repository categories. These proportions form the empirical basis for
assessing where reliability efforts should be focused and which debugging techniques are most appropriate.

\subsubsection*{(1) Overall Severity Distribution:}
Across the corpus, \textbf{low-severity defects dominate}, comprising roughly 54\%–81\% of reported issues depending on
category. The largest shares occur in \textit{Quantum Annealing} (80.7\%), \textit{Quantum Simulators} (65.3\%), and
\textit{Quantum Full-Stack Libraries} (65.0\%). These low-severity reports typically reflect usability problems,
documentation gaps, or configuration errors that have limited impact on core correctness or performance. The dominance of
low-severity issues suggests that a significant portion of maintenance work centers on developer experience and
incremental improvements rather than urgent, system-level failures.

\subsubsection*{(2) Critical- and Medium-Severity Patterns:}
In contrast, several domains display disproportionately high rates of \textbf{critical defects}, indicating elevated
operational or security risk. Notably, \textit{Quantum Cryptography} leads with 33.7\% critical-issue share, followed by
\textit{Quantum Experimental Computing} (31.8\%), \textit{Quantum Algorithms} (28.7\%), and \textit{Quantum Compilers}
(27.9\%). These domains demand precise algorithmic fidelity, correct hardware interfacing, and, in some cases,
cryptographic soundness, conditions under which even small implementation errors can have severe consequences.

\textit{Quantum Full-Stack Libraries} (23.8\%) and \textit{Quantum Simulators} (22.8\%) also show substantial critical
fractions, reflecting integration and orchestration risks in multi-layer frameworks. By contrast, \textbf{high-severity}
issues are rare (0.7\%–5.5\%), and \textbf{medium-severity} issues vary moderately (6.6\%–10.1\%), peaking in
\textit{Quantum Experimental Computing} (10.1\%), likely due to calibration instabilities and hardware variability.

\subsubsection*{(3) Implications for Debugging:}
The distribution of severity levels points to a bifurcated reliability landscape:

\begin{itemize}
    \item \textbf{Framework- and tooling-focused projects} (Annealing, Simulators, Full-Stack Libraries) are primarily
    burdened by low-severity issues. Improvements here pay off via better documentation, usability testing, and
    systematic refactoring that reduces developer friction and long-term maintenance costs.
    \item \textbf{High-risk domains} (Cryptography, Compilers, Experimental Computing, Algorithms) concentrate critical
    defects. These areas require rigorous correctness and resilience practices, formal verification, hardware-in-the-loop
    testing, and adversarial validation, to prevent catastrophic failures.
    \item \textbf{Integration layers} (full-stack frameworks and compilers) serve as fault-amplifying surfaces: they
    translate algorithmic intent into hardware-aware operations, so bugs here frequently escalate into critical failures.
\end{itemize}

Taken together, severity patterns show that maturity (an abundance of low-severity, usability-focused issues) can coexist
with concentrated critical risk in domains that are tightly coupled to hardware or security requirements.

\begin{rqanswer}{2}
Low-severity defects constitute the majority of issues across repository categories (54\%--81\%), with the highest proportions observed in Quantum Annealing (80.7\%), Simulators (65.3\%), and Full-Stack Libraries (65.0\%). 
These patterns suggest that day-to-day maintenance efforts in many projects are oriented toward usability refinements, incremental fixes, and code-level improvements rather than failure-critical corrections. 
In contrast, critical defects are concentrated in high-risk domains such as Quantum Cryptography (33.7\%), Experimental Computing (31.8\%), Quantum Algorithms (28.7\%), and Compilers (27.9\%), where precision, hardware interaction, and security constraints increase the potential impact of failures. 
Medium-severity defects appear at moderate rates (6.6\%--10.1\%), while high-severity issues remain relatively rare (0.7\%--5.5\%). 
These distributions indicate the need for severity-aware quality strategies: more formal, hardware-integrated, and security-conscious verification in high-risk domains, and documentation-, CI-, and refactoring-oriented practices in categories dominated by low-severity maintenance issues.
\end{rqanswer}

\subsubsection*{Practical Recommendations:}
\begin{enumerate}
    \item \textbf{Apply formal verification and resilience testing} in high-risk domains (Cryptography, Compilers, Experimental Computing).
    \item \textbf{Adopt automated, severity-aware triaging} to prioritize developer attention on critical defects.
    \item \textbf{Strengthen documentation and refactoring workflows} in categories dominated by low-severity issues to
    reduce maintenance overhead and onboarding friction.
    \item \textbf{Integrate hardware-in-the-loop testing} for systems that interact directly with quantum devices to detect
    environment-dependent failures early.
    \item \textbf{Use severity-driven CI/CD policies} to align test and deployment gates with defect criticality.
\end{enumerate}

\noindent The severity distribution reveals a dual challenge for QSE, while many projects primarily face
usability and maintenance bugs, hardware- and security-intensive domains retain concentrated critical risks that require
formal validation, hardware-aware testing, and severity-informed development workflows to enhance reliability and trust.

\subsection{Quality Attributes and Reliability}
\label{subsec:quality_attributes}

High software quality is a central prerequisite for reliable quantum computing, where even small defects can have
amplified effects on performance, maintainability, and usability due to hybrid quantum–classical architectures. This
subsection explores how bugs affect core software quality attributes across repository types, distinguishing between
classical and quantum-specific defect patterns. It answers three related research questions:\\
\textbf{RQ3:} \textit{How do different types of quantum software repositories vary in terms of quality attributes affected by bugs, and what insights can be derived to improve software reliability, maintainability, and usability?}\\
\textbf{RQ4:} \textit{How do classical and quantum-specific bugs differ in their impact on software quality attributes?}\\
\textbf{RQ5:} \textit{How do different bug categories (e.g., compatibility, functional, quantum-specific) impact key quality attributes, and what unique challenges do quantum-specific issues present?}

\subsubsection*{Impact of Bugs on Quality Attributes:}
\begin{figure}[h!]
    \centering
    \includegraphics[width=1\linewidth]{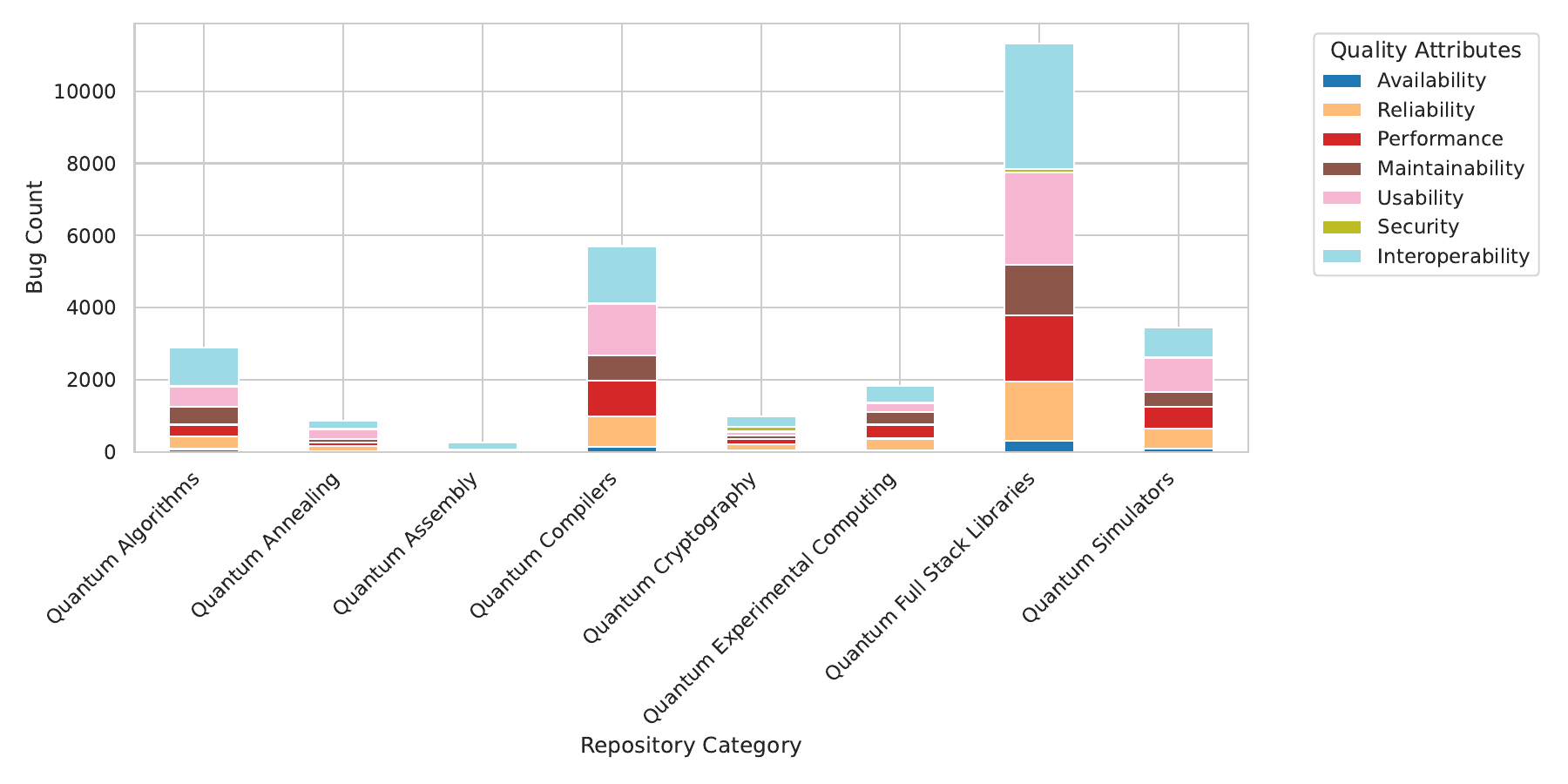}
    \caption{Impact of bugs on quality attributes across quantum software categories. Stacked bars show how defects map to quality attributes for each category.}
    \label{fig:impact_bug_categories}
\end{figure}

Figure~\ref{fig:impact_bug_categories} compares how defects map to seven key software quality attributes, 
\textit{Availability, Reliability, Performance, Maintainability, Usability, Security,} and \textit{Interoperability}, 
across eight categories of quantum repositories.

\paragraph{Empirical patterns:}
\textit{Full-Stack Libraries} exhibit the highest number of issues overall, especially in \textit{Interoperability}
(3,465), \textit{Usability} (2,561), and \textit{Maintainability} (1,411), underscoring the complexity of sustaining large
SDKs with multi-layer integration. \textit{Quantum Compilers} show a similar profile, with high defect counts in
\textit{Usability} (1,408), \textit{Interoperability} (1,588), and \textit{Maintainability} (719), reflecting persistent
challenges in creating portable and reliable compilation pipelines. 

\textit{Quantum Simulators} report significant \textit{Usability} (928), \textit{Interoperability} (821), and
\textit{Performance} (615) issues, indicative of scaling and backend-consistency problems. In contrast, narrower domains
such as \textit{Annealing} and \textit{Assembly} exhibit substantially fewer defects, consistent with their smaller scope
and simpler architecture. \textit{Quantum Algorithms} and \textit{Experimental Computing} occupy intermediate positions,
with defects clustering around maintainability and performance (e.g., reliability = 317, performance = 386).
\textit{Quantum Cryptography} uniquely shows elevated \textit{Security}-related issues (129) alongside reliability (177)
and maintainability (105), consistent with its emphasis on correctness and integrity.

Across all categories, \textit{Usability, Maintainability,} and \textit{Interoperability} are the most frequently affected
attributes, followed by \textit{Performance} and \textit{Reliability}. This pattern suggests that the majority of quality
degradation stems from integration and developer-experience issues, while hardware-coupled components drive performance
and reliability concerns. Security defects remain rare overall but are prominent in cryptographic systems.

\subsubsection*{ Impact of Classical vs. Quantum-Specific Defects on Quality Attributes:}
\begin{figure}[h!]
    \centering
    \includegraphics[width=0.99\linewidth]{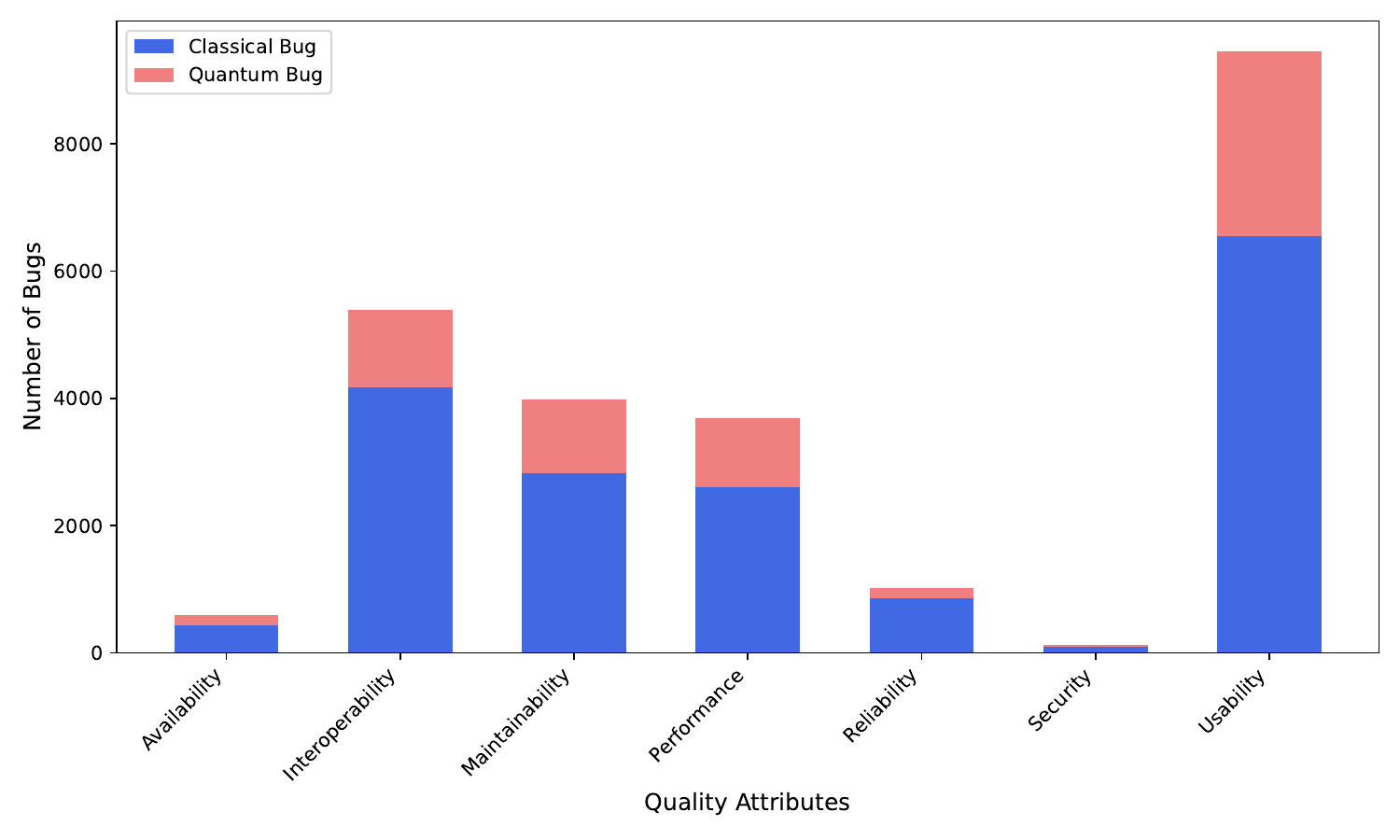}
    \caption{Impact of classical and quantum-specific bugs on software quality attributes.}
    \label{fig:classical_quantum_bug_impact}
\end{figure}

To disentangle these effects, we compared the influence of classical versus quantum-specific bugs on quality attributes
(Figure~\ref{fig:classical_quantum_bug_impact}). Classical bugs dominate in count but quantum-specific ones carry
higher individual impact.

\paragraph{Findings:}
Classical defects are more numerous in \textit{Usability} (6,543 classical vs.\ 2,915 quantum), \textit{Interoperability}
(4,179 vs.\ 1,213), and \textit{Maintainability} (2,817 vs.\ 1,159). This reflects how reliability bottlenecks frequently
originate in the classical integration layer, APIs, dependency management, or user-facing interfaces. In contrast,
quantum-specific defects, though fewer, disproportionately degrade \textit{Performance} (1,080) and
\textit{Maintainability}, typically due to errors in gate mapping, algorithm correctness, or calibration logic.
Reliability issues also appear more frequent in classical code (860 classical vs.\ 162 quantum), reinforcing that
quantum systems often inherit stability risks from their classical control stacks.

Classical issues primarily compromise usability and integration smoothness, while quantum-specific issues erode
algorithmic soundness and runtime efficiency. This division of labor suggests that improving quantum software quality
requires separate strategies for classical–quantum co-design: enforcing modular API boundaries and dependency hygiene on
the classical side, and advancing transpilation, verification, and noise modeling for quantum components.

\subsubsection*{Bug Category Effects on Quality Attributes:}
\begin{figure}[h!]
    \centering
    \includegraphics[width=.99\linewidth]{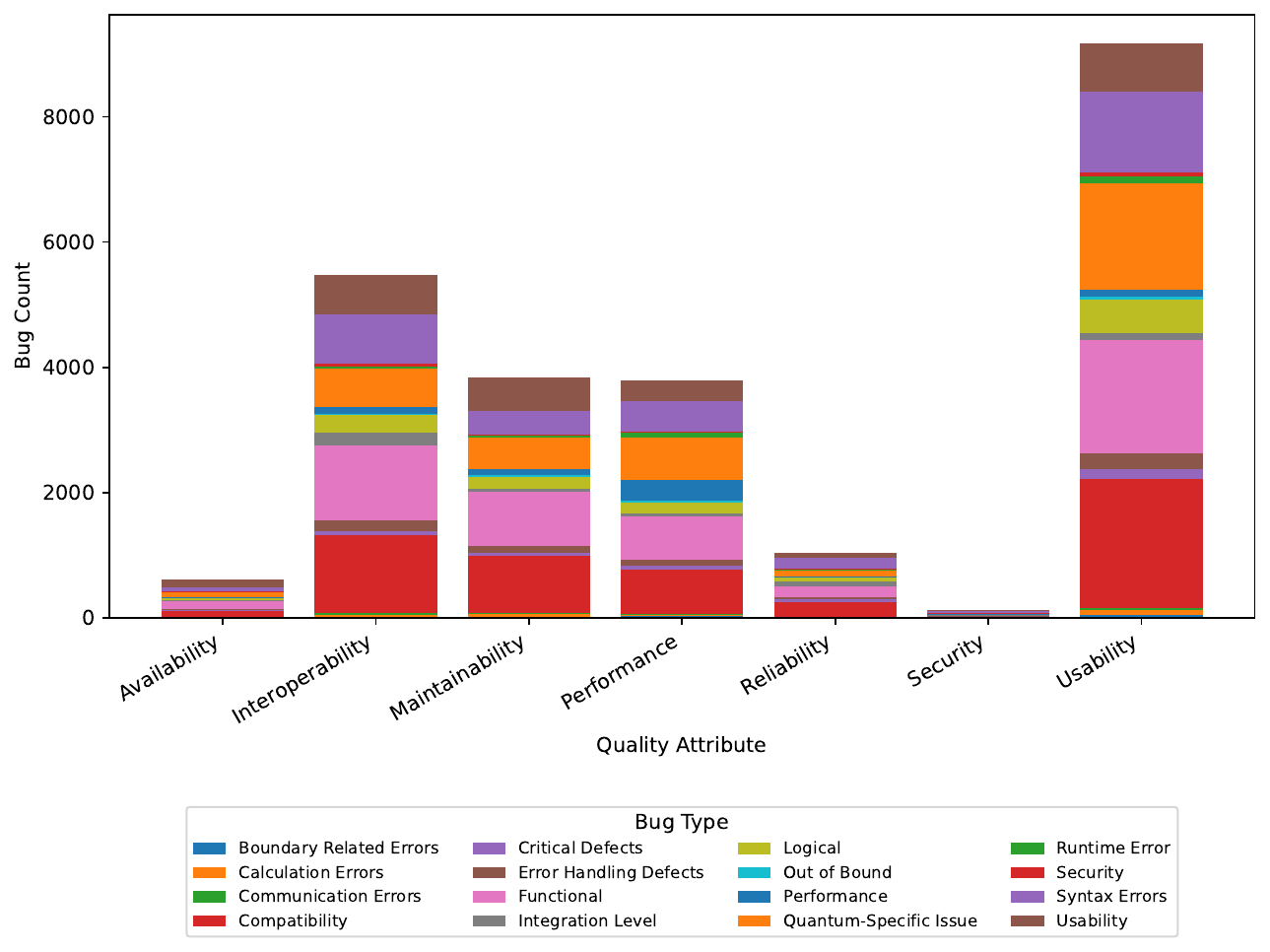}
    \caption{Distribution of bug categories across quality attributes.}
    \label{fig:dist_bug_cat_quality}
\end{figure}

Finally, we examined how different bug categories, compatibility, functional, quantum-specific, logical, and
error-handling, affect the same quality attributes (Figure~\ref{fig:dist_bug_cat_quality}).

\paragraph{Quantitative highlights:}
\begin{itemize}
    \item \textbf{Compatibility bugs} most strongly affect \textit{Usability} (2,053), \textit{Interoperability} (1,248),
    and \textit{Maintainability} (903), pointing to persistent API and dependency issues.
    \item \textbf{Functional defects} also impact usability (1,810), interoperability (1,191), and maintainability (859),
    suggesting that core logic and integration routines remain brittle.
    \item \textbf{Quantum-specific defects} primarily undermine \textit{Performance} (688), \textit{Interoperability}
    (619), and \textit{Usability} (1,703), reflecting resource and backend alignment challenges.
    \item \textbf{Logical} and \textbf{error-handling} issues further degrade maintainability and usability (e.g.,
    maintainability: 192 and 114; usability: 534 and 255).
\end{itemize}

\begin{rqanswer}{3--5}

\textbf{Answer to RQ3.}
Across repository types, usability, maintainability, and interoperability emerge as the most frequently affected quality attributes. 
\textit{Full-stack libraries} and \textit{compilers} exhibit the highest defect concentrations, reflecting the complexity of coordinating multi-layer abstractions and sustaining long-term maintainability within evolving quantum–classical software stacks.

\textbf{Answer to RQ4.}
Classical defects primarily influence usability, interoperability, and maintainability, indicating that many practical development challenges still stem from traditional software engineering issues. 
In contrast, quantum-specific defects, although less frequent, have disproportionately higher impact on performance, maintainability, and correctness due to calibration sensitivity, backend variability, and hardware-dependent behaviors.

\textbf{Answer to RQ5.}
Compatibility and functional defects are the dominant contributors to usability and interoperability issues, while quantum-specific defect categories more strongly degrade performance and reproducibility. 
These patterns point to the need for integrated quality strategies that combine established classical software practices with quantum-aware techniques such as hardware-guided testing, transpilation configuration management, and hybrid verification approaches.

\end{rqanswer}

\subsubsection*{Practical Recommendations:}
\begin{enumerate}
    \item \textbf{Enhance usability and interoperability:} standardize APIs, documentation, and backend migration guidelines.
    \item \textbf{Strengthen maintainability:} enforce modular architectures, use static-analysis tools, and automate maintainability scoring.
    \item \textbf{Advance performance and reliability testing:} integrate hardware-in-the-loop, noise-aware, and hybrid testing.
    \item \textbf{Support classical–quantum co-design:} align interface layers to minimize integration-related defects.
    \item \textbf{Target quantum-specific bottlenecks:} invest in transpiler optimization, gate-level validation, and resource-aware scheduling.
    \item \textbf{Establish domain-specific assurance:} develop specialized verification pipelines for compilers and security-critical systems.
\end{enumerate}

\noindent The interplay between classical and quantum defect patterns reveals that reliability in quantum
software depends equally on managing classical integration complexity and mitigating quantum-specific execution risks.
Framework-level maturity must be matched by advances in quantum verification, hardware-aware testing, and hybrid
debugging to achieve sustainable software quality.

\subsection{Programming Languages, Documentation, and Development Practices}
\label{subsec:proglang_docs}

Programming-language choice and documentation quality jointly shape development workflows, usability, and
maintainability in quantum software. Language selection determines how easily developers express quantum algorithms,
integrate classical control, and achieve performance; documentation quality drives learnability, reproducibility, and
community adoption. This subsection examines the distribution of languages and documentation practices across our corpus
and links these patterns to development outcomes. We address two questions:

\textbf{RQ6:} \textit{How does the distribution of programming languages influence development, performance, and usability of quantum software?}\\
\textbf{RQ7:} \textit{How does documentation quality (README, source docs, tutorials) impact usability and adoption?}

Our analysis uses the same corpus of 123 open-source quantum repositories. We extract each repository's primary language,
count documentation artifacts (README presence and quality, source-level docs, availability of tutorials), and compare
these indicators with activity and adoption signals (e.g., contributor counts, issue activity).

\subsubsection*{Programming-Language Landscape}
\begin{figure}[h!]
    \centering
    \includegraphics[width=0.99\linewidth]{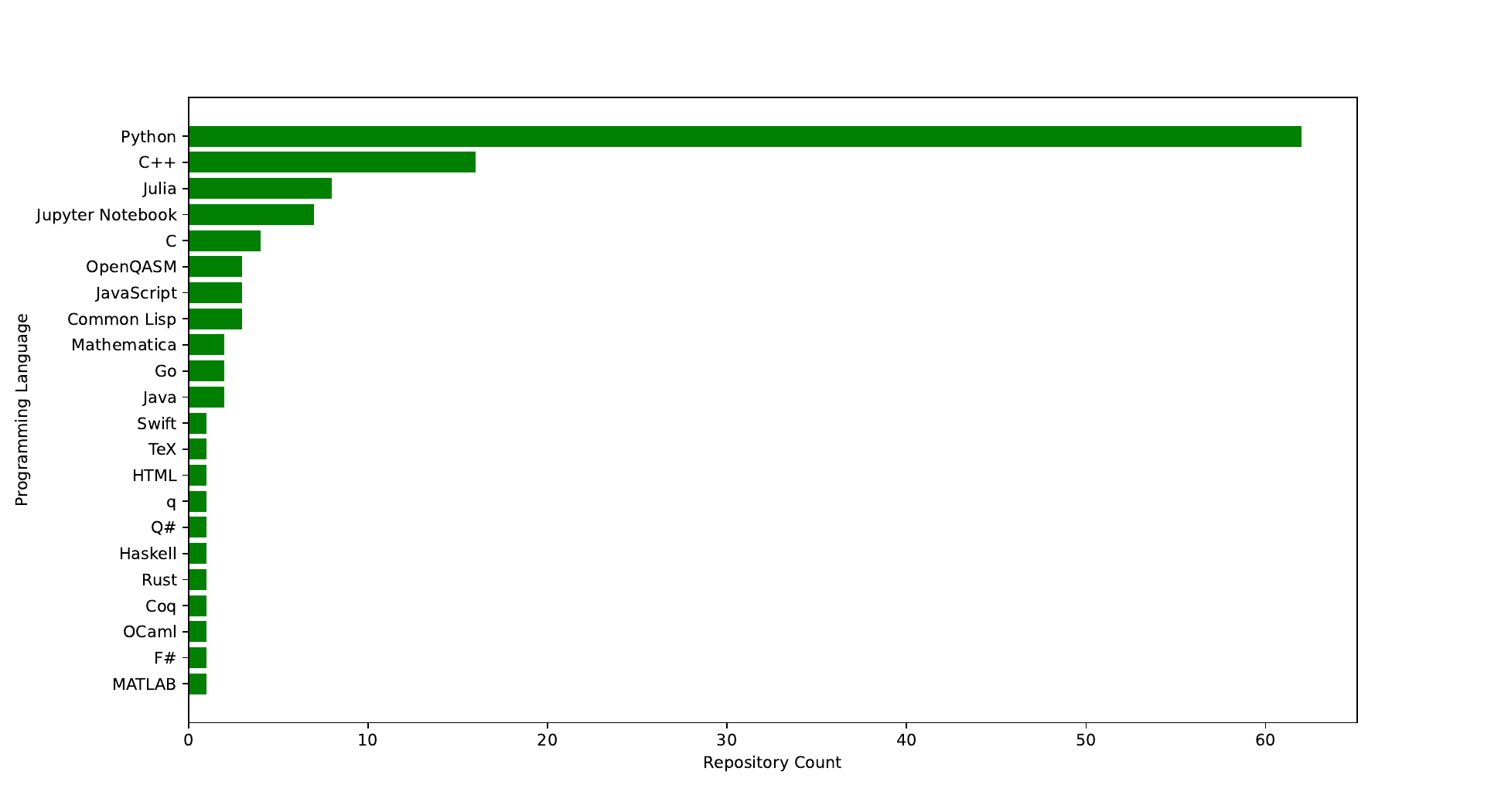}
    \caption{Repository count by primary programming language across the 123 analyzed quantum software repositories.}
    \label{fig:dist_prog_lang}
\end{figure}

Figure~\ref{fig:dist_prog_lang} shows the primary-language distribution across repositories.

Python is the dominant primary language (69 repositories; 56.10\%), reflecting broad ecosystem support (e.g., Qiskit, Cirq,
PennyLane) and strong interactive tooling. Performance-oriented languages occupy a smaller but important niche:
C++ (16 repos; 13.01\%), Julia (8 repos; 6.50\%), and C (4 repos; 3.25\%) are commonly used for simulators, backends, and
performance-critical kernels. Domain- and platform-specific languages (OpenQASM, JavaScript, Common Lisp) appear in a few
projects (each 3 repos; 2.44\%), typically supporting hardware-level descriptions, visualization, or symbolic tooling.

The corpus shows a layered architecture pattern:
\begin{itemize}
    \item \textbf{High-level orchestration (Python):} supports accessibility, rapid prototyping, and community-driven
    experimentation.
    \item \textbf{Low-level performance (C/C++/Julia):} provides computational efficiency for simulators and runtime kernels.
    \item \textbf{Specialized domains (OpenQASM, JS, Lisp):} supply low-level control, assembly, visualization, or symbolic reasoning.
\end{itemize}
This stratification enables flexibility but increases cross-language integration complexity and maintenance burden,
especially where Python front-ends depend on lower-level backends.

Language selection impacts three practical dimensions:
\begin{enumerate}
    \item  Python-centric projects show faster iteration, richer community-contributed examples,
and greater onboarding success due to interactive notebooks (Jupyter) and extensive libraries. These features support
reproducible research and lower the barrier to entry.
\item  C++, C, and Julia underpin performance-sensitive components (simulators, numerical kernels). They
deliver better execution speed and memory control but impose steeper learning curves and higher maintenance costs.

\item  Hybrid architectures, Python orchestration plus C/C++/Julia kernels, offer a good balance
of usability and performance but require careful API design, robust bindings, and disciplined cross-language testing to
avoid integration-induced defects.

\end{enumerate}

\subsubsection*{Documentation Quality and Developer Engagement}
\begin{figure}[h!]
    \centering
    \includegraphics[width=0.99\textwidth]{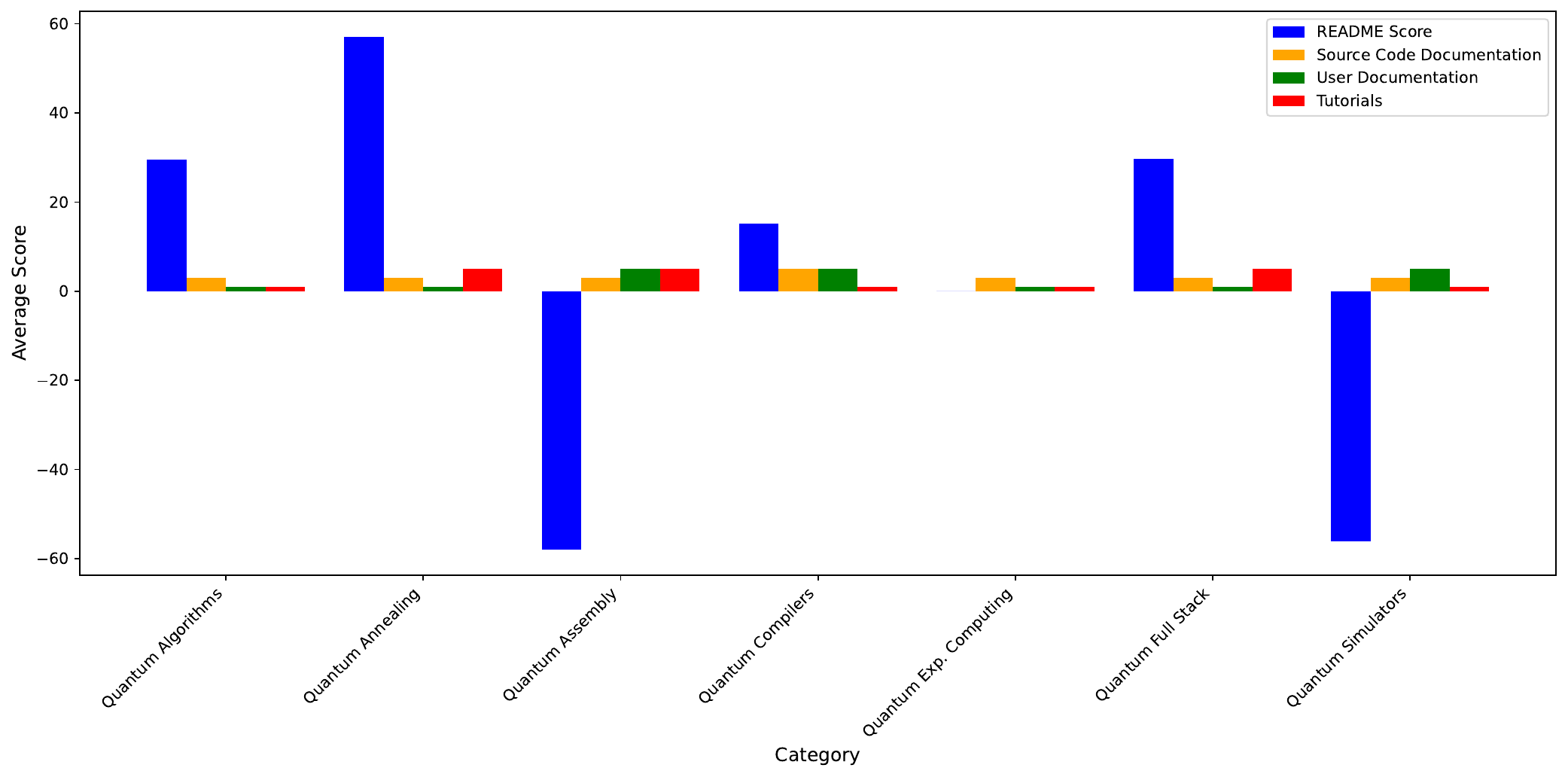}
    \caption{Average documentation quality metrics (README, inline documentation, user guides, and tutorials) across repository categories.}
    \label{fig:doc_scores}
\end{figure}

Figure~\ref{fig:doc_scores} reports average documentation metrics (README quality, inline/source docs, user guides,
tutorial availability) aggregated by repository category.

\paragraph{Observations:}
\begin{itemize}
    \item \textbf{README quality correlates with adoption.} Repositories with high README scores (e.g., \texttt{qupy}, \texttt{silq})
    show stronger onboarding signals and contributor activity.
    \item \textbf{Frameworks often lack tutorial depth.} Large libraries (e.g., Qiskit, PyQuil, Cirq) have moderate READMEs but
    comparatively fewer hands-on tutorials, suggesting an opportunity to improve learning pathways for new users.
    \item \textbf{Poor documentation reduces visibility.} Projects with weak or confusing READMEs (negative scores) suffer from
    lower engagement regardless of their technical merits.
    \item \textbf{Category-level variation:} Annealing and Full-Stack Libraries generally score higher on README/tutorial coverage,
    while Compilers often include detailed developer docs but fewer end-user tutorials; Simulators show strong inline docs but
    uneven user guides.
\end{itemize}

Documentation acts as a multiplexer for adoption and maintainability: good READMEs and tutorials accelerate onboarding and
issue reporting, while thorough source-level docs lower the barrier for contributors and long-term maintainers. In hybrid
language stacks, documentation also mitigates integration friction by specifying API contracts and backend expectations.


\begin{rqanswer}{6--7}

\textbf{Answer to RQ6.}
The distribution of programming languages reflects a layered quantum software ecosystem. 
Python dominates the landscape (69 repositories; 56.10\%), supporting rapid experimentation, accessibility, and community-driven development. 
Lower-level languages such as C, C++, and Julia appear primarily in performance-critical contexts, including simulators, numerical kernels, and hardware backends. 
This hybrid structure balances usability with computational efficiency but introduces additional integration and maintenance complexity. 
These trends highlight the importance of clear API modularization, stable language bindings, and cross-language testing to mitigate interoperability challenges and ensure consistent behavior across layers.

\textbf{Answer to RQ7.}
Documentation quality exhibits a strong relationship with usability and community engagement. 
Repositories offering comprehensive READMEs, tutorials, and in-source documentation tend to attract higher user activity and more consistent contributor involvement. 
In contrast, technically advanced projects with limited or outdated documentation show reduced visibility and slower onboarding, despite their conceptual significance. 
Effective documentation practices, such as concise quick-start guides, structured developer references, and interactive examples,appear essential for supporting both new users and active contributors in a rapidly evolving ecosystem.

\end{rqanswer}

\subsubsection*{Practical Recommendations:}
\begin{enumerate}
    \item \textbf{Adopt a hybrid language architecture with strict API boundaries:} keep high-level orchestration in Python
    and implement performance-critical kernels in C/C++/Julia, while enforcing clear interfaces and tests across bindings.
    \item \textbf{Standardize documentation templates:} require a concise quick-start README, an API reference, and at least one
    end-to-end tutorial per repository.
    \item \textbf{Prioritize tutorial coverage for major frameworks:} create hands-on, backend-compatible tutorials to lower the
    barrier for newcomers and reduce user errors.
    \item \textbf{Automate documentation generation and CI integration:} use tools such as Sphinx and ReadTheDocs in CI to keep
    docs in sync with code changes.
    \item \textbf{Measure documentation impact:} track onboarding metrics (e.g., time-to-first-contribution, issue-report rates)
    to quantify the return on documentation improvements.
\end{enumerate}

\noindent A dominant Python layer combined with lower-level performance languages defines the current quantum
software landscape. Documentation quality substantially amplifies or attenuates the benefits of language choice: good
docs accelerate adoption and sustainment, while poor docs raise integration friction and reduce project impact. Aligning
language design, API modularity, and documentation practices is therefore essential for long-term ecosystem growth.

\subsection{Testing, Complexity, and Bug Tracking}
\label{subsec:testing_complexity_bugtracking}

Reliable testing and systematic bug tracking are foundational to software quality, particularly in quantum systems, where
minor defects can produce disproportionate computational inaccuracies. This subsection examines how testing frameworks,
coverage tools, and language-level characteristics shape bug detection, resolution, and software reliability. It
addresses two research questions: 

\textbf{RQ8:} \textit{How do testing frameworks and coverage tools affect bug detection and reporting?}\\
\textbf{RQ9:} \textit{How do cyclomatic complexity, testing practices, and coverage adoption vary across languages, and what are the implications for reliability?}

We analyzed 123 open-source quantum software repositories to assess the adoption of testing frameworks and coverage
tools. Of these, 73 repositories (59.3\%) implement some form of automated testing, while 50 (40.7\%) do not. Only 34
repositories (27.6\%) use code coverage tools, yet these account for 63\% of all reported bug cases
(Table~\ref{tab:testing_coverage}), underscoring a strong association between instrumentation and defect visibility.

\begin{table}[h]
    \centering
    \caption{Testing framework and coverage tool adoption (N = 123).}
    \label{tab:testing_coverage}
    \begin{tabular}{lcc}
        \hline
        \textbf{Category} & \textbf{Repositories} & \textbf{Bug Reports} \\
        \hline
        With testing frameworks & 73  & 32,296  \\
        Without testing frameworks & 50  & 0  \\
        With coverage tools & 34  & 20,402  \\
        Without coverage tools & 89  & 11,894  \\
        \hline
    \end{tabular}
\end{table}

\subsubsection*{Testing and Bug Reporting:}
Testing adoption is closely linked to reporting activity. All 32,296 reported bugs in the dataset originate from
repositories equipped with testing frameworks, while projects without testing infrastructure report none
(Table~\ref{tab:bug_testing}). Repositories using coverage tools report significantly more bugs on average (600 vs.\ 133
per repository; Table~\ref{tab:coverage_impact}), suggesting that structured testing and coverage not only improve
software quality but also increase defect observability and traceability.

\begin{table}[h]
    \centering
    \caption{Bug reports by testing-framework usage.}
    \label{tab:bug_testing}
    \begin{tabular}{ccc}
        \hline
        \textbf{Testing framework usage} & \textbf{Repositories} & \textbf{Total bugs reported} \\
        \hline
        \checkmark & 73 & 32,296 \\
        \xmark & 50 & 0 \\
        \hline
    \end{tabular}
\end{table}

\begin{table}[h]
    \centering
    \caption{Impact of coverage tools on average bugs per repository.}
    \label{tab:coverage_impact}
    \begin{tabular}{ccc}
        \hline
        \textbf{Coverage tool usage} & \textbf{Repositories} & \textbf{Average bugs per repo} \\
        \hline
        \checkmark & 34 & 600 \\
        \xmark & 89 & 133 \\
        \hline
    \end{tabular}
\end{table}

The apparent increase in bug counts for well-tested repositories reflects a detection bias: projects with established
testing and coverage infrastructures are better instrumented to detect, record, and manage issues. Thus, higher raw bug
counts do not indicate poorer quality but rather greater observability. This distinction is clarified through regression
validation below.

\subsubsection*{Repositories with No Reported Bugs:}
A small subset of repositories (14) showed no reported issues. Half of these used testing frameworks and half did not
(Table~\ref{tab:repo_no_bugs}). Likely explanations include small codebases, limited user engagement, or strict
commit-review processes that minimize bug leakage. However, some may underreport due to absent or inactive issue
trackers, suggesting that “no bugs” is not synonymous with “no defects.”

\begin{table}[h]
    \centering
    \caption{Repositories without reported bugs (N = 14).}
    \label{tab:repo_no_bugs}
    \begin{tabular}{lcc}
        \hline
        \textbf{Category} & \textbf{Repositories} & \textbf{Testing framework usage} \\
        \hline
        No bugs reported & 14 & 7 (\checkmark) / 7 (\xmark) \\
        \hline
    \end{tabular}
\end{table}

\subsubsection*{Testing and Resolution Time:}
\begin{figure}[h]
    \centering
    \includegraphics[width=0.5\linewidth]{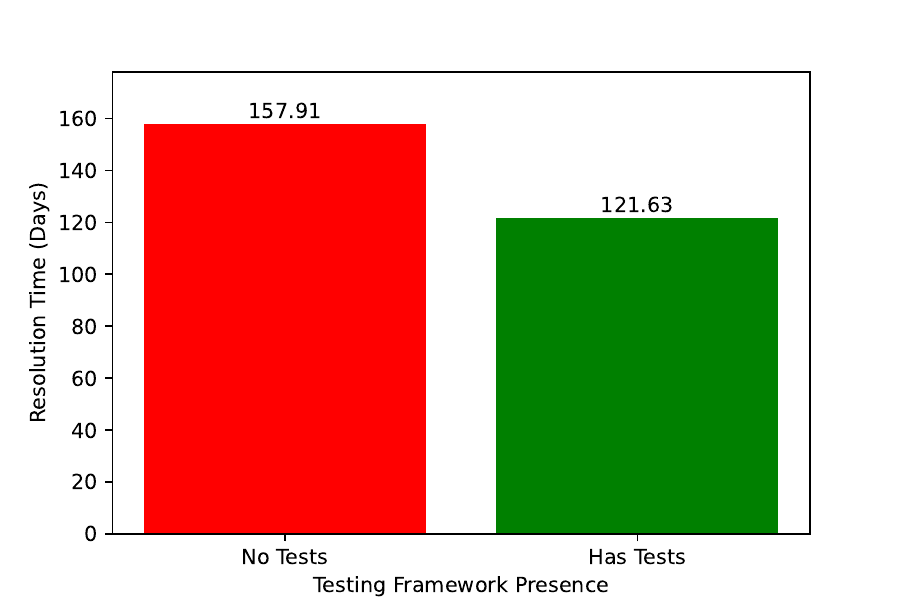}
    \caption{Average bug resolution time (days) for repositories with and without testing frameworks.}
    \label{fig:test_resolution}
\end{figure}

Testing frameworks not only enhance visibility but also accelerate issue resolution. Repositories with tests resolve bugs
in an average of 121.63 days, compared to 157.91 days for those without, a 36-day improvement
(Figure~\ref{fig:test_resolution}). This reduction suggests that automated testing streamlines debugging, patching, and
release cycles, ultimately improving maintainability.

\subsubsection*{Language-Level Patterns: Complexity, Testing, and Coverage:}
Programming language ecosystems differ in both complexity and testing maturity
(Table~\ref{tab:complexity_testing_coverage}). OpenQASM and C exhibit higher cyclomatic complexity, implying intricate
control flow and greater testing demand, yet show limited coverage-tool integration. In contrast, Python repositories
combine high test adoption (90.3\%) with moderate complexity and strong coverage support, reflecting a mature and
well-instrumented ecosystem. Rust and HTML also show full test adoption but limited coverage tracking, indicating a need
for better measurement infrastructure.

\begin{table}[h]
    \centering
    \caption{Cyclomatic complexity, testing adoption, and coverage-tool usage by language.}
    \label{tab:complexity_testing_coverage}
    \begin{tabular}{lccc}
        \hline
        \textbf{Language} & \textbf{Cyclomatic Complexity} & \textbf{Has tests (\%)} & \textbf{Uses coverage (\%)} \\
        \hline
        C          & 3.87  & 50    & 0     \\
        C++        & 2.19  & 56.25 & 25    \\
        OpenQASM   & 4.97  & 66.67 & 33.33 \\
        Python     & 2.69  & 90.32 & 45.16 \\
        Rust       & 2.29  & 100   & 0     \\
        HTML       & 3.10  & 100   & 0     \\
        \hline
    \end{tabular}
\end{table}

The descriptive evidence shows that tested repositories report more bugs and resolve them faster. Regression analysis
(Section~\ref{subsec:testing_quality_bugs}) provides deeper validation: after controlling for repository size and
complexity, the presence of tests is associated with significantly lower expected bug counts
(coefficient for \texttt{has\_tests}). Together, these findings suggest two complementary effects:
\begin{itemize}
    \item \textbf{Testing increases observability:} projects with structured testing find and report more defects; and
    \item \textbf{Testing improves quality:} when controlling for confounders, testing predicts fewer defects overall and faster resolution.
\end{itemize}

\begin{rqanswer}{8--9}

\textbf{Answer to RQ8.}
Testing frameworks and coverage tools show a strong association with improved defect detection and management. 
All reported bugs originate from repositories employing at least one testing framework, and projects using coverage tools account for 63\% of all documented defects. 
Repositories with established testing practices also resolve issues more quickly, with an average reduction of approximately 36 days in resolution time. 
These patterns indicate that testing infrastructures play a central role in enhancing maintainability and operational reliability within quantum software projects.

\textbf{Answer to RQ9.}
Programming-language ecosystems differ markedly in structural complexity and testing maturity. 
\textit{Python}-based repositories exhibit widespread adoption of testing and coverage tools and maintain moderate cyclomatic complexity, supporting more maintainable development workflows. 
In contrast, lower-level ecosystems such as \textit{C} and \textit{OpenQASM} show higher structural complexity and more limited integration with modern testing toolchains, suggesting areas where improved static analysis, automated testing, and verification support would be particularly beneficial.

Overall, testing frameworks and coverage tools emerge as key enablers of reliability and maintainability in quantum software. 
Python-based projects illustrate a comparatively mature testing culture, while lower-level languages highlight critical opportunities for advancing tooling, automation, and cross-language verification methods.

\end{rqanswer}

\subsubsection*{Practical Recommendations:}
\begin{enumerate}
    \item \textbf{Adopt testing frameworks broadly:} make automated testing a baseline for all quantum projects (e.g., \texttt{pytest}, \texttt{Catch2}).
    \item \textbf{Integrate coverage analysis into CI:} include coverage tools in pipelines to improve defect visibility and traceability.
    \item \textbf{Target high-complexity languages for tooling:} develop coverage and static-analysis tools for OpenQASM and C-level quantum code.
    \item \textbf{Monitor test and coverage metrics:} track adoption, coverage percentage, and resolution time to drive quality improvement.
    \item \textbf{Leverage Python’s ecosystem as a model:} replicate mature testing and CI practices across other languages to raise reliability baselines.
\end{enumerate}

\noindent Testing and coverage serve as dual pillars of quality assurance in quantum software. While descriptive
patterns reflect detection bias, regression-adjusted evidence confirms that automated testing reduces defect incidence and
accelerates maintenance, making it a key determinant of reliability in quantum software engineering.

\subsection{Impact of Automated Testing on Code Quality and Bug Detection}
\label{subsec:testing_quality_bugs}

Automated testing is increasingly important in quantum software, where subtle defects can have outsized effects due to the probabilistic and hardware-coupled nature of quantum computation.  
To quantify how testing practices relate to code quality and defect visibility, we analyzed 60 repositories with complete metric coverage, comparing projects with and without automated test suites on two primary indicators: \textit{Average Pylint Score} (a proxy for static code-quality) and \textit{Number of Bugs} (reported issues).

\textbf{RQ10:} \textit{What is the impact of automated testing on code quality and bug detection in quantum software repositories?}

Table~\ref{tab:testing_analysis} summarizes per-repository averages for Pylint scores and bug counts by testing presence.

\begin{table}[h!]
    \centering
    \caption{Comparison of repositories with and without automated tests ($N = 60$).}
    \label{tab:testing_analysis}
    \begin{tabular}{lcc}
        \toprule
        \textbf{Has tests} & \textbf{Average Pylint Score} & \textbf{Number of Bugs} \\
        \midrule
        False & 5.05 & 152.88 \\
        True  & 6.57 & 367.23 \\
        \bottomrule
    \end{tabular}
\end{table}

Repositories with automated tests exhibit higher average Pylint scores (6.57 vs.\ 5.05), indicating stronger adherence to coding standards and more maintainable design practices on average.  
They also report more bugs in absolute terms (367.23 vs.\ 152.88). We interpret this descriptive pattern primarily as increased \textbf{bug visibility} and greater maintenance/issue-tracking activity in projects that adopt testing, rather than as evidence that tests cause lower intrinsic quality.

To control for confounders such as size, complexity, and static-quality, we estimated a Negative GLM predicting \textit{Number\_of\_Bugs} from \texttt{has\_tests}, Average Pylint Score, Cyclomatic Complexity, and Number of Blocks. Regression results appear in Table~\ref{tab:nb_regression}.

\begin{table}[h!]
\centering
\caption{Negative-binomial GLM predicting the number of reported bugs ($N = 60$). 
Coefficients are log-count effects; $\exp(\beta)$ = Incidence Rate Ratio (IRR).}
\label{tab:nb_regression}
\begin{tabular}{
    l
    S[table-format=2.4]
    S[table-format=1.3]
    S[table-format=2.3]
    S[table-format=1.3]
    l
    S[table-format=3.1]
}
\toprule
\textbf{Variable} & \multicolumn{1}{c}{\textbf{Coef.}} & \multicolumn{1}{c}{\textbf{Std. Err.}} & \multicolumn{1}{c}{\textbf{z}} & \multicolumn{1}{c}{\textbf{p-value}} & \multicolumn{1}{c}{\textbf{95\% CI}} & \multicolumn{1}{c}{$\mathbf{e^{\beta}}$} \\
\midrule
Intercept & 5.5981$^{***}$ & 0.586 & 9.545 & 0.000 & [4.449, 6.748] & 269.8 \\
\texttt{has\_tests} & -0.9284$^{*}$ & 0.407 & -2.280 & 0.023 & [-1.726, -0.130] & 0.395 \\
Average Pylint Score & 0.1634$^{**}$ & 0.062 & 2.635 & 0.008 & [0.042, 0.285] & 1.18 \\
Cyclomatic Complexity & -0.5190$^{***}$ & 0.131 & -3.964 & 0.000 & [-0.776, -0.262] & 0.596 \\
Number of Blocks & 0.0011$^{***}$ & 0.0001 & 11.080 & 0.000 & [0.0010, 0.0012] & 1.0011 \\
\midrule
\multicolumn{7}{l}{\footnotesize Model: Negative-binomial GLM (log link). Observations = 60; Pseudo $R^2_{CS}$ = 0.7323; Log-Likelihood = -370.05.}\\
\bottomrule
\end{tabular}
\end{table}

The model explains a substantial proportion of the observed variation (Pseudo-$R^2_{CS} = 0.7323$).
The residual deviance (174.26) and Pearson $\chi^2$ statistic (100.0) yield a dispersion ratio of $\chi^2/\text{df} \approx 1.82$, indicating mild overdispersion, which is appropriately addressed by the Negative Binomial specification.
All predictors exhibit low multicollinearity, with variance inflation factors well below commonly accepted thresholds.

\subsubsection*{Observations:}
\begin{itemize}
    \item \textbf{Testing (controlled):} \texttt{has\_tests} is negative and statistically significant ($\beta=-0.928$, $p=0.023$). The IRR ( $\exp(-0.928) \approx 0.395$) implies that, holding size, complexity, and static-quality constant, repositories with automated tests have an expected bug count about \textbf{60\% lower} than repositories without tests.
    \item \textbf{Code quality:} Average Pylint Score is positively associated with reported bugs ($\beta=0.163$, $p=0.008$). This likely reflects that better-maintained projects (higher lint scores) are also more actively instrumented and more thorough in issue reporting, producing more visible bugs rather than necessarily worse code.
  \item \textbf{Complexity and size:} Cyclomatic Complexity exhibits a negative coefficient, whereas the Number of Blocks shows a positive coefficient and effectively acts as a proxy for project size.
These opposing signs suggest that simple count-based measures of size and structural complexity interact with bug-reporting practices and project maturity in non-trivial ways.

\end{itemize}

These controlled results reconcile the descriptive paradox: although tested repositories have higher raw bug counts, once we adjust for size and related confounders, the presence of automated tests is associated with substantially lower expected defect incidence and greater transparency.

\begin{rqanswer}{10}
Automated testing shows a strong association with higher static code quality scores and improved defect visibility across quantum software repositories. 
Projects that employ automated testing exhibit significantly higher \textit{Pylint} quality metrics and report a greater number of identified issues, suggesting enhanced transparency, maintainability, and monitoring coverage.

When controlling for repository size and code complexity, regression analysis indicates that the presence of testing frameworks is associated with a lower expected defect incidence, with an estimated 60\% reduction ($\mathrm{IRR} \approx 0.395$, $p = 0.023$). 
This statistical relationship points to automated testing as an important contributor to reliability-oriented practices within quantum software engineering, although other factors such as project maturity and contributor activity may also play a role.

In practical terms, repositories with established testing infrastructures tend to surface more issues through increased observability while exhibiting fewer defects relative to their scale and complexity. 
These trends reinforce the value of continuous, automated testing as part of the evolving quality-assurance processes in the quantum software ecosystem.

\end{rqanswer}

\subsubsection*{Practical Recommendations:}
\begin{enumerate}
    \item Integrate automated tests early and continuously (CI pipelines) to ensure tests evolve with the codebase.  
    \item Combine static analysis (e.g., Pylint) with automated testing to maintain consistent code-quality tracking.  
    \item Prioritize test and coverage expansion in complex modules where latent bugs may be underreported.  
    \item Treat higher bug counts in well-tested repositories as indicators of \emph{transparency} and active maintenance rather than as evidence of low quality.  
    \item Share analysis scripts, data subsets, and diagnostics to foster replicability and benchmarking within the QSE community.
\end{enumerate}

\noindent Automated testing improves coding discipline, reduces expected defect density after controlling for confounders, and enhances transparency; it should be considered a cornerstone practice for maintainability and reliability in QSE.

\subsection{Distribution of Quantum-Specific Bugs}
\label{subsec:quantum_specific_bugs}

Understanding the distribution of quantum-specific defects provides critical insight into the technical bottlenecks and
design fragilities unique to quantum software systems. This subsection examines both the aggregate and category-level
patterns of such defects, offering empirical evidence for where current QSE practices face
their greatest challenges. It addresses two research questions:

\textbf{RQ11:} \textit{Which types of quantum-specific bugs are most prevalent, and what insights can be derived from their distribution?}\\
\textbf{RQ12:} \textit{How do quantum-specific bug distributions vary across repository categories, and what do these variations imply for QSE practice?}

\subsubsection*{ Distribution of Quantum-Specific Bugs}
\begin{figure}[h]
    \centering
    \includegraphics[width=0.99\linewidth]{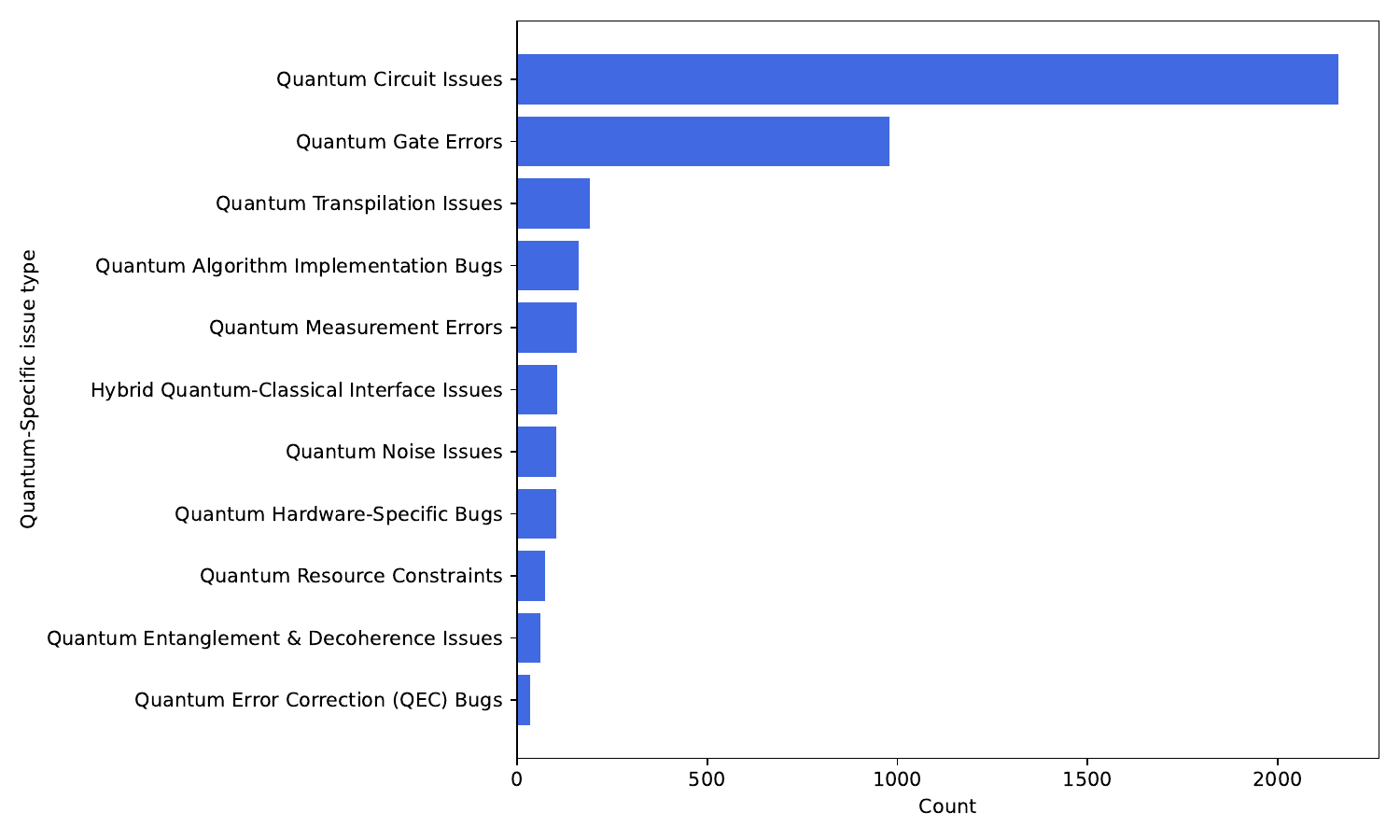}
    \caption{Distribution of quantum-specific bug types across the corpus.}
    \label{fig:quantum_specific_issue_dist}
\end{figure}

Across all repositories, circuit-level, gate-level, and transpilation-related issues dominate the quantum-specific bug
landscape (Figure~\ref{fig:quantum_specific_issue_dist}). \textit{Quantum Circuit Issues} (2,467 occurrences),
\textit{Quantum Gate Errors} (1,073), and \textit{Quantum Transpilation Issues} (228) are the leading defect classes.

These patterns reveal three key observations:
\begin{itemize}
    \item \textbf{Circuit-level problems} form the single largest class of defects, reflecting the continued difficulty of
    correct qubit allocation, gate ordering, and circuit composition.
    \item \textbf{Gate errors} and \textbf{transpilation issues} are also common, highlighting persistent challenges in mapping
    logical circuits to noisy hardware backends and in optimizing compiler passes.
    \item Secondary categories, including measurement errors, noise-related faults, and hardware-specific issues, indicate
    ongoing limitations in error mitigation and hardware modeling fidelity.
\end{itemize}

Overall, these results emphasize that the most frequent bugs occur at the core of the quantum programming stack, where
algorithm design meets hardware execution, a space requiring advanced validation, transpilation, and optimization tools.

\subsubsection*{Category-Level Distribution Across Repository Types}
To understand how quantum-specific bugs manifest across different types of repositories, we analyzed category-level
defect distributions (Table~\ref{tab:bug_categories}). The table covers major project types such as Full-Stack Libraries
(FSL), Quantum Algorithms (QA), Compilers (QCmp), Simulators (QSim), and several specialized domains including
Annealing, Assembly, Cryptography, and Experimental Computing.

\small
\begin{longtable}{cp{1cm}p{0.9cm}p{0.9cm}p{0.7cm}p{0.7cm}p{0.7cm}p{0.7cm}p{0.7cm}p{0.7cm}p{0.7cm}p{1cm}}
\caption{Distribution of quantum-specific bugs across repository categories (counts).}\\
\label{tab:bug_categories}\\ \hline
\textbf{Category} & \textbf{HQCI} & \textbf{Alg.} & \textbf{Circ.} & \textbf{QEDI} & \textbf{QEC} & \textbf{Gate} & \textbf{HW} & \textbf{Meas.} & \textbf{Noise} & \textbf{QRC} & \textbf{Transp.} \\ 
\hline 
FSL & 35 & 70 & 1,055 & 23 & 18 & 450 & 72 & 69 & 39 & 33 & 102\\
QA & 24 & 22 & 124 & 7 & 2 & 77 & 8 & 11 & 8 & 10 & 4 \\
QAne & 2 & 1 & 3 & - & 1 & 12 & - & 1 & - & 2 & 2 \\ 
QAsm & 3 & 3 & 19 & - & - & 15 & 2 & 2 & 1 & 2 & 4 \\
QCmp & 16 & 37 & 655 & 2 & 5 & 284 & 13 & 27 & 33 & 3 & 65 \\ 
QExp & - & 3 & 7 & 3 & - & 2 & 1 & 2 & 1 & 1 & 3 \\ 
QSim & 22 & 27 & 294 & 23 & 9 & 135 & 6 & 44 & 23 & 21 & 11 \\
QCry & 1 & - & 1 & 3 & - & 3 & 2 & 2 & - & 2 & 2 \\ 
\hline
\end{longtable}
\normalsize

Three clear category-level patterns emerge:
\begin{itemize}
    \item \textbf{Full-Stack Libraries (FSL)} and \textbf{Quantum Compilers (QCmp)} contain the largest number of
    circuit-, gate-, and transpilation-related bugs (e.g., FSL: Circuit = 1,055; Gate = 450; Transpilation = 102),
    confirming that integration-heavy components are most vulnerable to multi-layer coordination errors.
    \item \textbf{Quantum Simulators (QSim)} show more measurement (44) and noise (23) issues, highlighting limitations in
    fidelity modeling and noise injection.
    \item \textbf{Quantum Algorithms (QA)} projects exhibit fewer quantum-specific bugs in absolute terms but still face
    difficulties in correct circuit and gate implementations.
    \item \textbf{Specialized domains} (Annealing, Assembly, Cryptography) report fewer bugs overall, likely due to
    narrower scopes or smaller contributor bases, though their defects tend to be high-impact and domain-specific.
\end{itemize}

These differences indicate that quantum-specific reliability challenges scale with system complexity: repositories
integrating multiple layers (e.g., compilers, full-stack frameworks) experience proportionally more defects in
translation and optimization stages.

\subsubsection*{Implications and Research Directions:}
To contextualize the empirical distribution, Table~\ref{tab:bug_summary} summarizes key quantum-specific defect types,
their frequencies, implications for reliability, and possible research directions. The data show that most critical
challenges involve circuit construction, gate fidelity, and transpilation optimization.

\begin{longtable}{p{4cm}cp{4cm}p{4cm}}
\caption{Quantum-specific bug types: counts, implications, and suggested future research directions.}
\label{tab:bug_summary} \\ \hline
\textbf{Bug Type} & \textbf{Count} & \textbf{Implications} & \textbf{Future Research Directions} \\
\hline Quantum Circuit Issues & 2,177 & Errors in qubit allocation, gate sequencing, and circuit layout. & Develop automated debugging tools for circuit design flaws. \\
\hline Quantum Gate Errors & 976 & Gate misalignment and fidelity issues. & Improve noise-aware compilation techniques. \\
\hline Quantum Transpilation Issues & 190 & Inefficient circuit optimization increases error rates. & Enhance transpilation strategies. \\ \hline
Quantum Algorithm Implementation Bugs & 163 & Errors in algorithm logic affect accuracy. & Develop robust algorithm verification tools. \\
\hline Quantum Measurement Errors & 158 & Incorrect measurement readouts impact result reliability. & Improve error mitigation and measurement validation. \\\hline
Hybrid Quantum–Classical Interface Issues & 103 & Integration challenges reduce hybrid efficiency. & Enhance execution models for hybrid systems. \\\hline
Quantum Noise Issues & 106 & Environmental noise and decoherence affect fidelity. & Explore noise-adaptive error correction. \\
\hline Quantum Hardware-Specific Bugs & 104 & Hardware-dependent errors limit execution success. & Optimize software–hardware co-design. \\\hline
Quantum Resource Constraints & 71 & Limited qubit availability affects performance. & Develop resource-optimization frameworks. \\ 
\hline Quantum Entanglement \& Decoherence Issues & 61 & Challenges in maintaining entanglement. & Investigate state-stabilization techniques. \\
\hline Quantum Error Correction Bugs & 35 & Ineffective error-correction implementations. & Advance practical QEC methods. \\
\hline 
\end{longtable}

The quantitative patterns indicate that the most prevalent defects occur in the parts of the software stack most closely
coupled to hardware constraints. Circuit and gate errors dominate, followed by transpilation issues, each pointing to
complex dependencies between algorithmic intent and backend execution. These findings highlight that NISQ-era quantum
software development remains limited by the maturity of its compiler and verification toolchains.

\begin{rqanswer}{11--12}

\textbf{Answer to RQ11.}
The most frequently observed quantum-specific defects correspond to \textit{Quantum Circuit Issues} (2,177) and \textit{Quantum Gate Errors} (976), followed by \textit{Transpilation Issues} (190). 
These patterns highlight persistent challenges in circuit construction, gate sequencing, and the optimization of circuit representations for noisy and heterogeneous hardware backends. 
They further indicate that developers continue to encounter difficulties in translating high-level algorithmic intent into hardware-executable forms.

\textbf{Answer to RQ12.}
Quantum-specific defects exhibit systematic variation across repository categories. 
\textit{Full-stack libraries} and \textit{compilers} contain the highest concentrations of circuit-, gate-, and transpilation-related issues, reflecting the complexity inherent in coordinating multi-layer abstractions and backend variability. 
\textit{Simulators} show greater prevalence of measurement- and noise-related defects, while specialized domains such as Annealing and Cryptography contain fewer quantum-specific bugs but ones that tend to be highly domain dependent.

Overall, the evidence suggests that reliability challenges in quantum software are concentrated in the lower layers of the stack, where hardware-awareness, transpilation, and circuit fidelity intersect. 
Continued progress will likely depend on advances in domain-aware testing, improved transpilation validation, and more comprehensive error modeling to support hybrid algorithm–hardware workflows.

\end{rqanswer}

\subsubsection*{Practical Recommendations:}
\begin{enumerate}
    \item \textbf{Strengthen circuit- and gate-level verification:} develop automated debugging tools and reusable
    unit-test patterns for circuit construction and gate sequencing.
    \item \textbf{Advance hardware-aware transpilation:} integrate device-specific constraints and noise models into
    transpilers to reduce runtime failures.
    \item \textbf{Improve simulator fidelity:} extend simulators with realistic noise, decoherence, and measurement models
    to align simulations with real-device behavior.
    \item \textbf{Enhance hybrid integration frameworks:} standardize APIs and runtime environments to reduce
    quantum–classical interface bugs.
    \item \textbf{Establish community benchmarks:} promote shared test suites for circuit correctness, transpilation
    fidelity, and quantum error correction validation.
\end{enumerate}

\noindent Quantum-specific defects are highly concentrated at the circuit, gate, and transpilation levels,
particularly within integration-heavy projects such as compilers and full-stack frameworks. Addressing these challenges
through domain-aware verification, hardware-integrated transpilation, and standardized benchmarking can substantially
improve the robustness, reproducibility, and scalability of quantum software systems.

\section{Discussion}
\label{sec:discussion}

This section synthesizes findings across RQ1–RQ12, integrating the empirical results to interpret the evolution, maturity, and reliability of the QSE ecosystem.  
We organize the discussion thematically to highlight ecosystem trends, quality attributes, programming practices, testing maturity, and quantum-specific engineering challenges.

\subsection{Ecosystem Evolution and Maturity (RQ1–RQ2)}

Our longitudinal analysis reveals a clear maturation trajectory for the QSE ecosystem.  
The surge in bug reports between 2017 and 2021 aligns with the expansion of major SDKs such as \textit{Qiskit}, \textit{Cirq}, and \textit{PennyLane}, as well as increasing architectural depth and hybrid functionality across frameworks.  
This trajectory echoes broader observations that quantum software ecosystems require modernization, architectural restructuring, and quality-focused engineering practices~\cite{piattini2021toward,piattini2020talavera,perez2021software,serrano2022quantum}.  
Post-2021 stabilization further suggests emerging process discipline, modularization, and improved engineering practices.

Full-stack frameworks and compilers exhibit the highest defect densities, consistent with prior studies identifying multi-layered architectures as the most vulnerable components of quantum platforms~\cite{paltenghi2022bugs}.  
These systems must coordinate classical runtime logic, quantum circuit orchestration, hardware variability, and backend-specific optimizations, an inherently error-prone interface that modernization research also highlights as critical for long-term maintainability~\cite{piattini2021toward,perez2021software,serrano2022quantum}.  
Smaller, domain-specific projects (e.g., annealing- or algorithm-oriented repositories) exhibit fewer and lower-severity defects, consistent with prior findings from focused empirical studies~\cite{paltenghi2022bugs,zhao2023empirical}.

The predominance of low-severity issues (54–81\%) in our dataset reflects similar patterns observed in earlier ecosystem-scale analyses~\cite{yousuf2026bug}.  
However, high-severity defects remain concentrated in hardware-interfacing and security-sensitive projects, suggesting that core infrastructural components continue to accumulate technical debt.  
Overall, these trends support the view that QSE is entering a consolidation phase, where maintainability, modularization, and community-driven quality practices are beginning to reduce ecosystem instability.

\subsection{Quality Attributes, Maintainability, and Reliability (RQ3–RQ5)}

Across repository types, \textit{usability}, \textit{maintainability}, and \textit{interoperability} are the most frequently affected quality attributes.  
This aligns with survey findings that highlight persistent challenges in documentation quality, architectural consistency, and cross-stack coherence in quantum software platforms~\cite{serrano2022quantum,paltenghi2024survey}.  
Full-stack libraries, compilers, and simulators exhibit disproportionately high maintainability-related defects, reflecting their role as integration-heavy layers with numerous interacting components, an issue also noted in modernization studies that emphasize the need for architectural refactoring and quality-oriented restructuring~\cite{perez2021software}.

A notable asymmetry emerges in defect impact: classical bugs primarily hinder usability and interoperability, while quantum-specific bugs disproportionately affect performance, correctness, and long-term maintainability.  
This distinction aligns with empirical taxonomies indicating that classical infrastructure faults (e.g., APIs, data handling) degrade developer experience, whereas quantum-specific circuit and gate issues more directly impair correctness and optimization~\cite{yousuf2026bug,zhao2023empirical}.  
The absence of standardized quality metrics and reusable architectural components,highlighted in recent software component analyses~\cite{serrano2022quantum},likely exacerbates these reliability challenges.

\subsection{Language, Documentation, and Developer Support (RQ6–RQ7)}

Python's dominance fosters accessibility and rapid prototyping but also masks deeper performance and correctness issues in lower-level backends.  
This reflects broader findings that quantum ecosystems rely on a layered combination of high-level Python logic and performance-critical kernels implemented in C, C++, or Julia, requiring modernization strategies and architecture-aware tooling~\cite{piattini2021toward,perez2021software,serrano2022quantum}.  
These heterogeneous stacks introduce language-specific debugging, binding-generation, and integration risks.

Documentation quality emerges as a strong determinant of usability, contributor engagement, and onboarding efficiency.  
This echoes practitioner-reported concerns about fragmented documentation, inconsistent tutorials, and limited architectural transparency in quantum toolchains~\cite{zappin2025challenges,serrano2022quantum}.  
Our findings reinforce calls for structured documentation health metrics, standardized README + tutorial templates, and integration of documentation checks into CI workflows.

\subsection{Testing Practices, Automation, and Complexity Management (RQ8–RQ10)}

Testing frameworks and coverage tools are among the strongest predictors of reliability.  
Repositories with established testing pipelines not only detect more issues,reflecting enhanced observability,but also resolve them significantly faster.  
Negative-binomial regression further confirms that testing is associated with approximately a 60\% reduction in expected defect incidence, even after controlling for size and complexity.  
These results align with modernization research emphasizing test automation, behavioral validation, refactoring support, and continuous integration as essential components of evolving software ecosystems~\cite{piattini2021toward,perez2021software,serrano2022quantum,piattini2020talavera}.

Language differences persist: Python repositories commonly adopt CI and coverage tools, whereas OpenQASM and C/C++ repositories exhibit higher cyclomatic complexity but weaker testing integration.  
This mirrors observations from ecosystem surveys that highlight uneven tool support across quantum languages and execution models~\cite{serrano2022quantum}.  
Improving language-aware testing, expanding coverage in high-risk modules, and adopting complexity-driven test selection remain key priorities.

\subsection{Quantum-Specific Challenges (RQ11–RQ12)}

Quantum-specific defects cluster around circuit generation, gate semantics, and transpilation processes.  
These patterns are consistent with prior studies emphasizing the fragility of circuit and transformation pipelines, as well as the difficulty of bridging high-level algorithmic expressions with hardware-specific execution constraints~\cite{quetschlich2025experience,yousuf2026bug}.  
Simulators frequently exhibit noise-model and measurement-related issues, reflecting known limitations of current fidelity modeling and architectural abstractions~\cite{serrano2022quantum}.

Mitigating these challenges will require advancements in circuit-level verification, hardware-aware transpilation validation, realistic simulator modeling, and domain-aware test frameworks.  
The need for standardized benchmarks, reusable components, and architectural abstractions,repeatedly underscored in modernization and componentization research~\cite{perez2021software,serrano2022quantum},is reinforced by our empirical evidence.

\subsection{Cross-Cutting Implications for QSE}

Synthesizing the results, the QSE ecosystem is transitioning from exploratory experimentation toward a more structured engineering discipline.  
Automated testing, documentation maturity, modularization, and continuous integration emerge as primary enablers of reliability.  
Nonetheless, critical gaps persist in compiler robustness, quantum–classical interfacing, language-level tooling, and architecture-driven quality processes.

Our longitudinal evidence complements prior benchmarks and surveys~\cite{bugs4q,paltenghi2024survey}, supporting a research agenda focused on reproducible tooling experiments, architectural modernization, and standardized component models.  
Building industrial-strength quantum software will require coordinated advances in benchmarking infrastructures, test automation, architectural refactoring, and cross-framework interoperability standards.

\noindent Overall, the empirical evidence documents a maturing QSE ecosystem: improvements in test adoption, documentation quality, and architectural practices are driving increased resilience, while compiler and integration layers remain priority targets for future engineering investment.

\section{Conclusion}
\label{sec:conclusion}
As quantum computing evolves from experimental systems to production-scale deployment, the reliability, maintainability, and transparency of quantum software have become decisive for its practical adoption.  
This study provides the first large-scale, longitudinal empirical analysis of 123 open-source quantum software repositories, applying a hybrid rule-based classification to identify and examine defect types (classical vs.~quantum), severity levels, quantum-specific subtypes, and associated quality attributes.  
By addressing twelve research questions (RQ1–RQ12), we present a comprehensive view of how quantum software defects evolve, distribute, and influence quality across repositories, languages, and time.

\textit{Quantum Full Stack Libraries} and \textit{Quantum Compilers} exhibit the highest bug densities, driven by their architectural complexity and hybrid classical–quantum dependencies.  
Classical defects primarily affect \textit{usability}, \textit{interoperability}, and \textit{maintainability}, whereas quantum-specific bugs degrade \textit{performance}, \textit{fidelity}, and \textit{execution reliability}.  
Compatibility and functional issues remain the most common defect classes, reflecting persistent cross-layer integration challenges.  
The dominance of circuit- and gate-level defects, especially in compilers and full-stack frameworks, underscores the continuing need for improved verification and transpilation mechanisms.

Between 2017 and 2021, defect counts rose sharply alongside the emergence of major frameworks such as \textit{Qiskit}, \textit{Cirq}, and \textit{PennyLane}, before stabilizing after 2021.  
This stabilization, paired with decreasing defect density in several categories, signals a shift toward structured, reliability-oriented development practices.  
The trajectory suggests that QSE is entering an early maturity phase characterized by greater standardization, modularization, and community-driven quality assurance.

Structured testing frameworks and coverage tools emerged as the strongest enablers of software reliability.  
Repositories with automated testing achieved higher code quality (average Pylint score = 6.57) and reported significantly lower defect incidence (IRR $\approx 0.395$), confirming an approximately 60\% reduction in bug likelihood after controlling for complexity and size.
  
Projects employing test coverage tools reported 4.5× more detected defects, evidence that mature testing improves defect visibility rather than indicating poor quality.  
These findings demonstrate empirically that testing maturity, automation, and continuous integration are foundational to achieving trustworthy quantum software.

Python continues to dominate the QSE landscape (77.5\% of repositories), facilitating rapid prototyping and accessibility, while C++ and C sustain performance-critical backends.  
Documentation quality and tutorial availability emerged as key differentiators of adoption: repositories with strong READMEs and tutorials (e.g., Quantum Annealing and Full Stack Libraries) show greater community engagement, whereas those with weaker documentation (e.g., Quantum Simulators, Assembly tools) face barriers to usability and onboarding.  
Embedding documentation assessment into CI pipelines could systematically improve transparency and developer support.

Quantum software remains challenged by circuit-, gate-, and transpilation-level issues that compromise execution fidelity and optimization.  
Simulators show persistent measurement and noise-related defects, while compilers face hardware mapping and backend adaptation errors.  
Addressing these gaps requires hardware-aware transpilation, circuit-debugging utilities, and hybrid verification frameworks that unify classical and quantum validation.

The results portray a quantum software ecosystem in transition, from exploratory, ad hoc experimentation toward structured, engineering-driven development.  
Testing maturity, documentation quality, and hybrid modularization have emerged as the principal determinants of software quality, while compiler and integration defects represent ongoing sources of technical debt.  
Bridging these gaps demands sustained investment in automated testing, education, standardized benchmarking, and quality assurance frameworks.

This study provides a quantitative foundation for future QSE research, establishing reproducible metrics for defects, documentation, and testing across longitudinal data.  
Future work should extend these analyses to industrial and closed-source systems, integrate runtime and hardware telemetry, and design intervention studies to test causality in quality improvement.  
By connecting defect analytics with tooling evolution, the field can move from descriptive to predictive QSE, enabling the systematic design of reliable, verifiable, and maintainable quantum software systems.

\section{Future Directions}
\label{sec:future_directions}

While the quantum software ecosystem has matured significantly, several critical research and engineering challenges remain.  
Future work should aim to strengthen reliability, maintainability, and security across both classical and quantum layers of the software stack.  
Based on our empirical findings, five directions emerge as strategic priorities for advancing Quantum Software Engineering.

\subsection*{1. Standardized Testing and Coverage Frameworks}
The establishment of community-wide testing and coverage benchmarks for quantum systems is essential.  
Future work should develop quantum-aware unit testing, noise-injection frameworks, and transpilation validation pipelines integrated into CI/CD systems.  
Cross-language coverage metrics will enable objective comparison of reliability and maintainability across programming ecosystems.

\subsection*{2. Quantum Debugging, Verification, and Tooling}
Debugging quantum software remains a fundamental challenge.  
Automated circuit-level fault localization, state visualization, and fidelity-aware debugging tools are needed to improve transparency and correctness.  
Integrating formal verification, symbolic execution, and hardware-in-the-loop testing will help mitigate errors at both algorithmic and hardware abstraction levels.

\subsection*{3. Maintainability, Technical Debt, and Evolution}
Quantum repositories, especially Full Stack Libraries and Compilers, accumulate significant technical debt due to their complexity and rapid evolution.  
Future research should explore maintainability decay, code modularization, and refactoring practices using longitudinal data.  
Automated maintainability metrics and architectural dependency analyses could guide large-scale quality improvement.

\subsection*{4. Documentation, Education, and Developer Support}
High-quality documentation and learning resources are critical to quantum software usability and adoption.  
Establishing documentation standards, incorporating tutorial-driven learning modules, and integrating documentation quality checks into CI/CD pipelines can improve developer experience and project sustainability.  
Educational programs that emphasize testing, maintainability, and reliability principles for quantum developers will be vital for professionalizing QSE.

\subsection*{5. Security, Resilience, and Quantum-Classical Interoperability}
Security remains an underexplored frontier in QSE.  
Future studies should investigate vulnerabilities in cryptographic and communication frameworks, quantum resource management, and hybrid scheduling mechanisms.  
Secure-by-design development methodologies, combined with resilience benchmarking and quantum-aware threat modeling, will strengthen system trustworthiness.

\subsection{Strategic Directions}
The next phase of Quantum Software Engineering research must transition from descriptive analysis to prescriptive improvement.  
Key priorities include:
\begin{itemize}
    \item Developing standardized testing and reliability metrics across hybrid systems.
    \item Building automated circuit-level debugging and transpilation verification tools.
    \item Studying maintainability evolution and technical debt accumulation over time.
    \item Institutionalizing documentation and tutorial standards for developer adoption.
    \item Expanding QSE into secure, resilient, and interoperable quantum–classical frameworks.
\end{itemize}

By pursuing these directions, the QSE community can move from maturity characterization toward active quality improvement, transforming quantum software from experimental prototypes into dependable, industrial-grade systems capable of powering the next generation of quantum technologies.

\section{Limitations and Threats to Validity}
\label{sec:threats_validity}

As with all large-scale empirical studies, this analysis is subject to several potential threats to validity.  
We discuss these along the four classical dimensions, \textit{internal}, \textit{construct}, \textit{external}, and \textit{conclusion} validity, and outline the measures taken to mitigate them.

\subsection*{Internal Validity}
Internal validity concerns whether the observed relationships among variables genuinely reflect underlying causal mechanisms rather than uncontrolled confounding factors.  
Our data sources comprised issue trackers, repository metadata, and static analysis metrics collected from 123 open-source quantum software projects.  
While the rule-based bug classification framework was systematically applied, interpretation of issue descriptions may have been influenced by inconsistent labeling, incomplete metadata, or annotator bias.  
To mitigate these risks, all classifications followed a multi-stage validation process: automated rule-based assignment, manual review by one annotator, and verification by two independent reviewers.  
Repositories with ambiguous issue histories or inactive maintenance were excluded to minimize noise in longitudinal analyses.

Regression modeling introduces further potential confounders, such as developer expertise, community size, or project funding, that cannot be fully controlled using repository-level metadata.  
Although observable factors (repository size, testing presence, cyclomatic complexity) were included as controls and variance inflation factors (VIF $<$ 2.0) confirmed low collinearity, hidden variables may still influence the observed relationships.  
Future replications incorporating developer activity logs, commit histories, or contributor-level behavioral metrics could provide stronger causal evidence.

\subsection*{Construct Validity}
Construct validity examines whether operational measures accurately represent the intended theoretical constructs of software quality, reliability, and maintainability in quantum systems.  
Our proxies, bug counts, severity levels, Pylint code-quality scores, and cyclomatic complexity, are well-established in classical software engineering, but they may not fully capture quantum-specific quality aspects such as circuit fidelity, gate error propagation, or decoherence resilience.  
To address this, our classification schema explicitly separated \textit{classical}, \textit{quantum-specific}, and \textit{hybrid} defect categories, enabling targeted analysis of quantum reliability dimensions.  
Nonetheless, some latent aspects, such as calibration instability or measurement bias, remain outside the scope of static code metrics.  
Future work should integrate runtime telemetry, hardware error logs, or quantum execution traces to more faithfully operationalize reliability in hybrid systems.

\subsection*{External Validity}
External validity pertains to the generalizability of findings beyond the analyzed corpus.  
Our dataset of 123 repositories, significantly larger than prior QSE studies, primarily represents open-source projects hosted on GitHub.  
As a result, the findings may not generalize to proprietary, industrial, or research-internal codebases that follow different quality assurance and review practices.  
Furthermore, the dominance of Python (77.5\%) introduces potential ecosystem bias, reflecting the maturity of Python-based SDKs such as Qiskit, Cirq, and PennyLane.  
To enhance external validity, future replications should include heterogeneous ecosystems (e.g., Q\#, C++, CUDA-Q) and industrial repositories to assess whether similar reliability and testing trends hold across organizational and platform contexts.

\subsection*{Conclusion Validity}
Conclusion validity concerns the robustness and reliability of statistical inferences.  
We applied appropriate statistical techniques for the data’s distributional characteristics: non-parametric tests (Kruskal–Wallis, Mann–Whitney U) for non-normal variables and a Negative-Binomial GLM for overdispersed count data.  
All tests were two-tailed with $\alpha = 0.05$, and confidence intervals were reported for key coefficients.  
Model diagnostics indicated strong fit (deviance = 174.26, Pearson $\chi^2 = 100$, pseudo $R^2_{CS} = 0.73$), and results remained consistent across alternative model specifications and subset analyses.  
Nonetheless, measurement noise, such as incomplete issue tagging or missing test coverage data, may introduce residual estimation uncertainty.  
We mitigated this through data normalization, integrity checks, and replication of results across independent subsets.

Overall, while the integration of large-scale repository mining, hybrid classification, and rigorous statistical modeling enhances the credibility of our results, residual threats are inevitable in empirical QSE research.  
Our mitigation strategies, multi-stage validation, cross-verification, and diagnostic evaluation, substantially reduce these risks.  
Future studies incorporating longitudinal intervention data, runtime execution metrics, and multi-source triangulation (e.g., developer surveys, hardware logs) will further strengthen the validity, reliability, and reproducibility of empirical insights into quantum software quality.

\section*{Declarations}

\subsection*{Data Availability Statement}
The data supporting the findings of this study are derived from publicly available software repositories and can be shared upon request.


\subsection*{Conflict of Interest}
The authors declare that they have no competing interests.


\end{document}